%% file: Article1_09_04.tex
\documentclass[
11pt, % Main document font size
a4paper, % Paper type, use 'letterpaper' for US Letter paper
oneside, % One page layout (no page indentation)
headinclude,
footinclude, % Extra spacing for the header and footer
]{article}
\input{structure.tex} % Include the structure.tex file which specified the document structure and layout

\begin{document}
%----------------------------------------------------------------------------------------
%	HEADERS
%----------------------------------------------------------------------------------------

\renewcommand{\sectionmark}[1]{\markright{\spacedlowsmallcaps{#1}}} % The header for all pages (oneside) or for even pages (twoside)
\renewcommand{\subsectionmark}[1]{\markright{\thesubsection~#1}} % Uncomment when using the twoside option - this modifies the header on odd pages
%\lehead{\mbox{\llap{\small\thepage\kern1em\color{halfgray} \vline}\color{halfgray}\hspace{0.5em}\rightmark\hfil}} % The header style

%\pagestyle{scrheadings} % Enable the headers specified in this block

\include{TitlePage}
%----------------------------------------------------------------------------------------
%	ABSTRACT
%----------------------------------------------------------------------------------------
%\section*{Abstract} % This section will not appear in the table of contents due to the star (\section*)
\include{Abstract}
%------------------ ----------------------------------------------------------------------
%	Significance Statement
%----------------------------------------------------------------------------------------
%\section*{Significance Statement} % This section will not appear in the table of contents due to the star (\section*)
%\include{SignificanceStatement}
\include{AuthorSummary}

%\newpage % Start the article content on the second page, remove this if you have a longer abstract that goes onto the second page

%\addtolength{\oddsidemargin}{-.875in}
%\addtolength{\evensidemargin}{-.875in}
%\addtolength{\textwidth}{6.75in}

%%----------------------------------------------------------------------------------------
%%	INTRODUCTION
%%----------------------------------------------------------------------------------------
\section{Introduction}
Human perception is a complex process which involves both sensory and cognitive components; it is thus sensitive to many factors other than the sensory input itself. 
As a result, responses to seemingly simple and controlled stimuli can be non-reproducible and fluctuate among repeated experiments. 
These fluctuations are inherent to Psychophysical performance; given the ubiquity of noise in the nervous system \cite{Faisal2008}, they could be plainly interpreted as such. 
However, recent studies suggest that they represent the influence on perception of factors other than the input: context~\cite{Sanabria2004,Wu2009}, history \cite{Howarth1956,Frund2014}, perceptual memory \cite{Magnussen1999}, attention\cite{Freiberg1937} and expectation~\cite{Raviv2012}.  
%Or that they serve some other (hidden) mechanisms in the perception process~\cite{Faisal2008}.
%Such influences have been documented in several sensory modalities under different conditions~\cite{Gilden1995}. 
\paragraph{} 
Temporal context and past history, both of the stimulus and of the response, influence perception~\cite{Gepshtein2005} with and without feedback on performance~\cite{Abrahamyan2016a,Raviv2012,Barack2016}.
Two opposing effects have been documented: 
On the one hand, observers tend to repeat previous responses, or to estimate signals as being similar to those previously perceived \cite{Howarth1956, Parducci1964, Lockhead1970, Anderson1971, Cross1973, Snyder2015}.
Such effects are predominant when stimuli are weak, near perceptual threshold
% responses to stimuli near the perceptual threshold display a positive trial-to-trial  correlation~
\cite{Parducci1964,Lockhead1970,Anderson1971}.  
On the other hand, a negative bias appears towards perceiving a signal as opposite or different from the previous ones. 
This effect is usually seen after exposure to a strong or sustained sensation, and can be manifested as an overshoot in the estimation of a new or different stimulus~\cite{Gibson1937,Chopin2012a}.

%These two effects can be understood in terms of the conflicting demands for stability and sensitivity imposed on the system by the external world. 
%Positive bias stabilizes perception near threshold against noise and ambiguity~\cite{Snyder2015}, taking advantage of the continuous and regular tendency of the natural world.
%In contrast, a negative bias contributes to maintaining sensitivity to new information and to the detection of changes~\cite{Clifford2000}.    
\paragraph{} 
In many psychophysical experiments, signals are presented in a random uncorrelated sequence~\cite{Holland1968,Cross1973,Ward1973,Luce1982,Gepshtein2005}. 
However even this simple design 
%One might have hoped that such a presentation would minimize response bias, but in fact the first descriptions of 
has revealed trial-to-trial response correlations~\cite{Verplanck1952,Blackwell1952,Pollack1954,Mcgill1957,Ward1973},
and was   
%The same experiment design was also 
used to study the biases and temporal context effects themselves~\cite{Holland1968,Cross1973,Ward1973,Luce1982,Gepshtein2005}. 
Most studied examined each trial relative to one or two previous inputs~\cite{Cross1973,Parducci1965,Raviv2012} or responses~\cite{Parducci1964},
or relative to a priming stimulus~\cite{Fischer2014,Snyder2015}. 
%However, even under random stimulus regimes, responses exhibit slow fluctuations and internal correlations \cite{Verplanck1952,Wertheimer1953,Gilden1995,Marom2011}. 
%Therefore it is of interest to design psychophysical stimuli that span a range of timescales and correlations, and to study the interplay of these timescales with those of internal processes participating in perc eption.

\paragraph{} 
Perception in a more natural setting entails continuously varying inputs that can be correlated in time; one may expect that history and context dependence of perception be coupled to these temporal structures. In the current study we explore this coupling %how sensory detection combines with history-dependent biases over multiple timescales, 
by manipulating input signals to have various temporal structures, and by employing dynamic analysis methods that highlight time-dependent aspects of the data. 
While the elementary detection task is seemingly simple, forming continuous streams of such tasks with various correlations and addressing the time dependencies of both input and output, reveal highly nontrivial facets of perception. 
It moreover brings the Psychophysical experiment a step closer to natural conditions.

\paragraph{} 
Our results show that in a near-threshold detection task, human observers exhibit simultaneously both positive and negative biases. Interestingly, these are characterized by different timescales: 
in the short term, responses are biased to be similar to previous ones. 
In the long term, a bias towards spontaneously switching the response is found. 
These two opposing biases are found to coexist in every stream of detection tasks we performed. However, the balance between them shifts by an interplay with the timescales of the input signal, to ultimately produce an outcome dependent on its temporal correlations.
%This effect is manifested, for example, by a probability of response alternation (POA) which is lower than a random sequence of independent detection events, for all types of input signals. 
%In the long term, a bias towards spontaneously switching the response is found. 
%This tendency can be seen as an exploratory behavior, and is emphasized when the stimulus varies slowly.
%The two opposing biases are manifested by a hysteresis of the psychometric curve conditioned on past input trend, which changes sign from positive at short times to negative at long times. 
%The interplay between the two biases and the timescale of the input signal results in a psychometric curve with markedly different slopes when averaged over the entire experiment. 
We construct a model of perception composed of two stages: a sensory stage, where the input signal is transformed into a detection probability; and a cognitive stage, where a final behavioral decision (detection / no detection) is made. 
By making both stages amenable to modification by history-dependent processes we can capture the entire set of experimental observations. 

%%----------------------------------------------------------------------------------------
%%	Experiment 
%%----------------------------------------------------------------------------------------
\section{Materials and Methods}

\subsection{Experiment Design}
\begin{figure}[H]
	\centering{
		\includegraphics[scale=1]{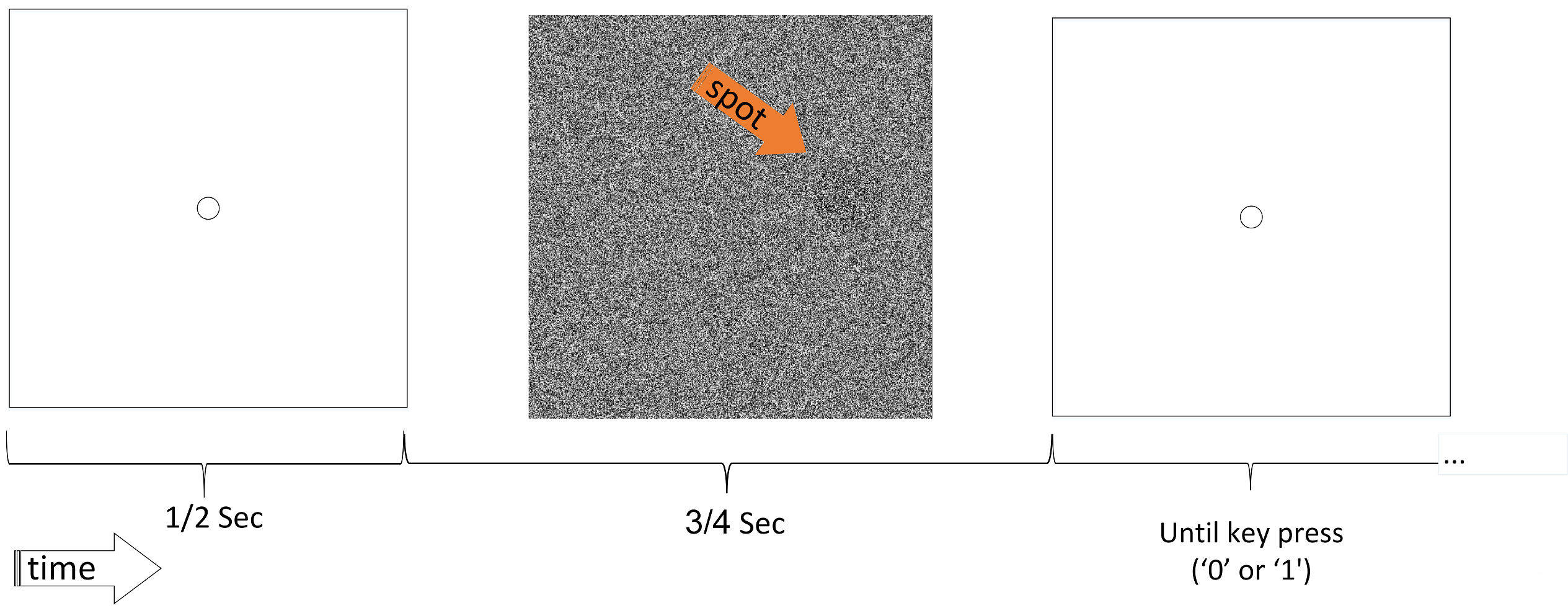}}
	\caption[]{\textbf{A Stimulus trial} starts with a blank screen with a fixation circle. 
		After 500ms the stimulus appears for 750ms,
		then the blank screen returns and remains until the observer reports whether or not a spot was detected.
The spot location, in this example with input level of 0.6, is marked here with an arrow for illustration.}
%\textbf{Stimulus trial}
	\label{fig:ExperiemntStim}
\end{figure}

\paragraph{}
\textit{ Elementary visual task.} The experiment contained a series of elementary visual detection tasks presented to human observers. The background screen showed a pattern of 500X500 black and white pixels, each drawn independently with probability 0.5 to be black or white. A new background with the ssame statistics was generated for each trial. On top of this background a darker circular spot, 60 pixels in diameter, was presented, 150 pixels from the center at a randomly chosen angle. The spot was composed of black and white pixels with probability larger than 0.5 for pixels to be black. This probability is referred to as the \textbf{input level}. In the example given in figure~\protect\ref{fig:ExperiemntStim} the input level was 0.6; this is well within the detectable range of most observers.
The input level varied from trial to trial.

Each trial began with a visual reset period of 500ms, in which a white screen with a fixation circle in the center was presented.
The stimulus was then presented for 750ms, after which the white screen returned (figure~\ref{fig:ExperiemntStim}). The observer was instructed to report with a key-press whether or not a spot was detected ('1' or '0' respectively). 
%In practice there was always a spot with probability for black pixels larger than 0.5, but sometimes it was too close to be detected. 
Pressing the response key immediately initialized the next trial.
The experiment was self-paced since response time had no upper time limit. 
Responses were accepted also during stimulus presentation without shortening the trial.

%\subsection{Temporal structure of stimulus}
\paragraph{}\textit{Temporal structure of stimulus.}
%Each experiment included 3 sessions of different correlative relations between consecutive stimulus.
%Thus, were created different contextual regimes which involve factors that are beyond the simple dependence in input level.
Input levels on consecutive trials were drawn from a normal distribution with fixed mean and variance, and with different temporal correlations: a ``White"  stimulus was created by drawing the level independently for each trial; the ``Pink" stimulus contained correlations, namely the changes in input levels among consecutive trials varied slowly; and finally the ``Brown" stimulus changed even more slowly in time. Examples of 500 input levels for each case are shown in Fig. 2 (top panels). These three types of temporal structures can also be characterized by their Power Spectral Density (PSD), shown in the lower panels of Fig. 2.
The Pink stimulus has a PSD decreasing as $\sfrac{1}{f}$, while the Brown decreases as $\sfrac{1}{f^{2}}$. 

\begin{figure}[H]
\centering
\includegraphics[scale=1,trim = 0.5cm 0mm 0.5cm 0mm, clip]{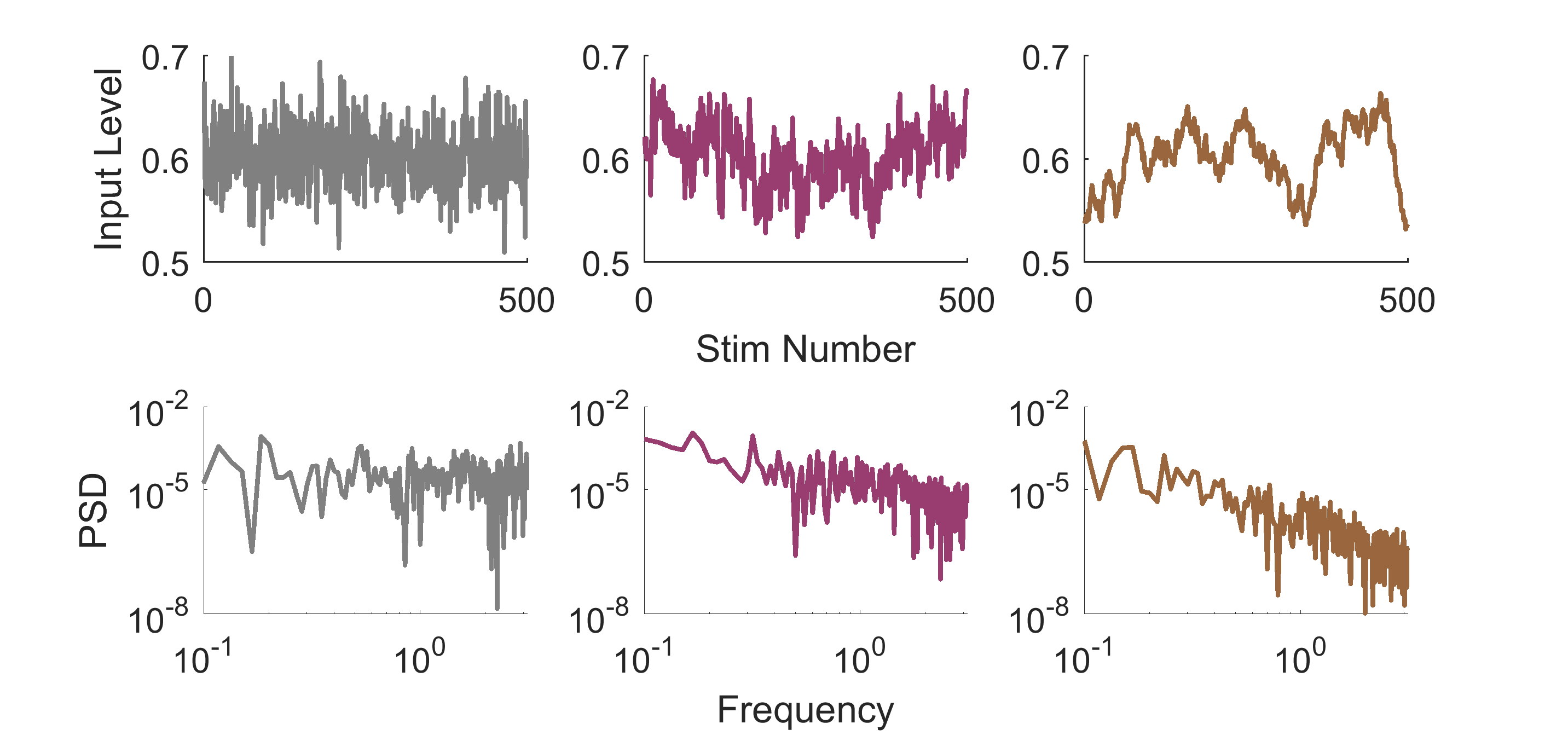}
\caption[]{\textbf{Temporal structure of stimuli.} 
	%from subject DM\_F\_24. 
%Each session is composed of 500 trials. 
Examples of input levels in their temporal order (top panels), and their corresponding power spectral densities (PSD, bottom panels), for the three types of stimuli used in the experiments. 
In the ``White" stimulus (left; grey lines), consecutive input levels are independent and the PSD is flat. 
In the ``Pink" stimulus (center; pink lines), consecutive input levels are correlated and the PSD decreases with frequency.  The ``Brown" session (right; brown lines) varies extremely slowly, consecutive input levels are highly correlated and the PSD decreases sharply.}
%\textbf{Experiment sessions}
\end{figure}

\begin{figure}[H]
	\centering
	\includegraphics[scale=1]{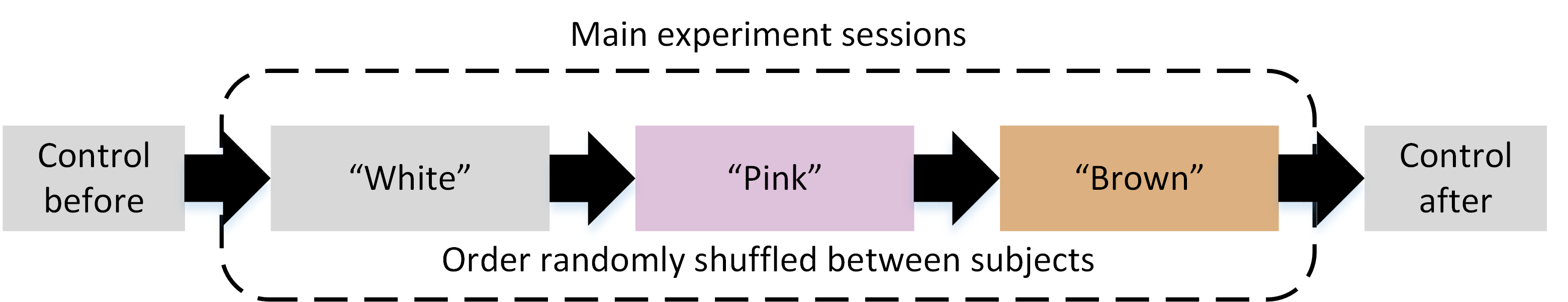}
	\caption[]{\textbf{Experiment structure} }
	\label{ExperimentStructure}
	%\textbf{Experimental protocol}
\end{figure}

%\subsection{Experimental protocol}
\paragraph{}\textit{Experimental protocol.} The entire experiment consisted of 3 main sessions, containing 500 consecutive trials each, with the three temporal structures described above. The sessions were presented in random order. Two additional control sessions were conducted before and after the main sessions  (figure~\protect\ref{ExperimentStructure}).

%%\subsection{Determination of input statistics of a subject}
\paragraph{}The first control session consisted of 100 trials, with stimulus levels independently drawn from a Gaussian distribution 
with mean 0.595 and standard deviation (STD) 
%0.0297, which is 
5\% of the mean. 
Responses were characterized by a psychometric curve (see next section for details). 
Analysis of this session was used to adjust the average input level in the following sessions to be 
at the individual threshold, i.e. the 50\% detection level, and the standard deviation to be 5\% of this value. 
This ensured a fair comparison among observers, with the stimulus levels being very close to the individual detection threshold throughout the experiment.

18 subjects aged 23-31, 9 females and 9 males, participated in the experiment.
They had regular or corrected-to-regular vision and were not diagnosed as having attention deficit disorders.
Two female subjects were excluded from the experiment for having extremely high positive responses ($>40\%$) for trials of very low input levels, which implied low credibility.
The experiments were conducted in a dark room where observers sat alone in front of a computer screen. No feedback on the tasks was provided along the experiment.
All participants signed consent forms, were paid for their time, and were naive to the purpose of the experiment.

\subsection{Statistical Analysis - Extracting psychometric curve parameters}
\label{subsec:Psychometric_curve}
\begin{figure}[H]
	\centering
	\includegraphics[scale=0.6]{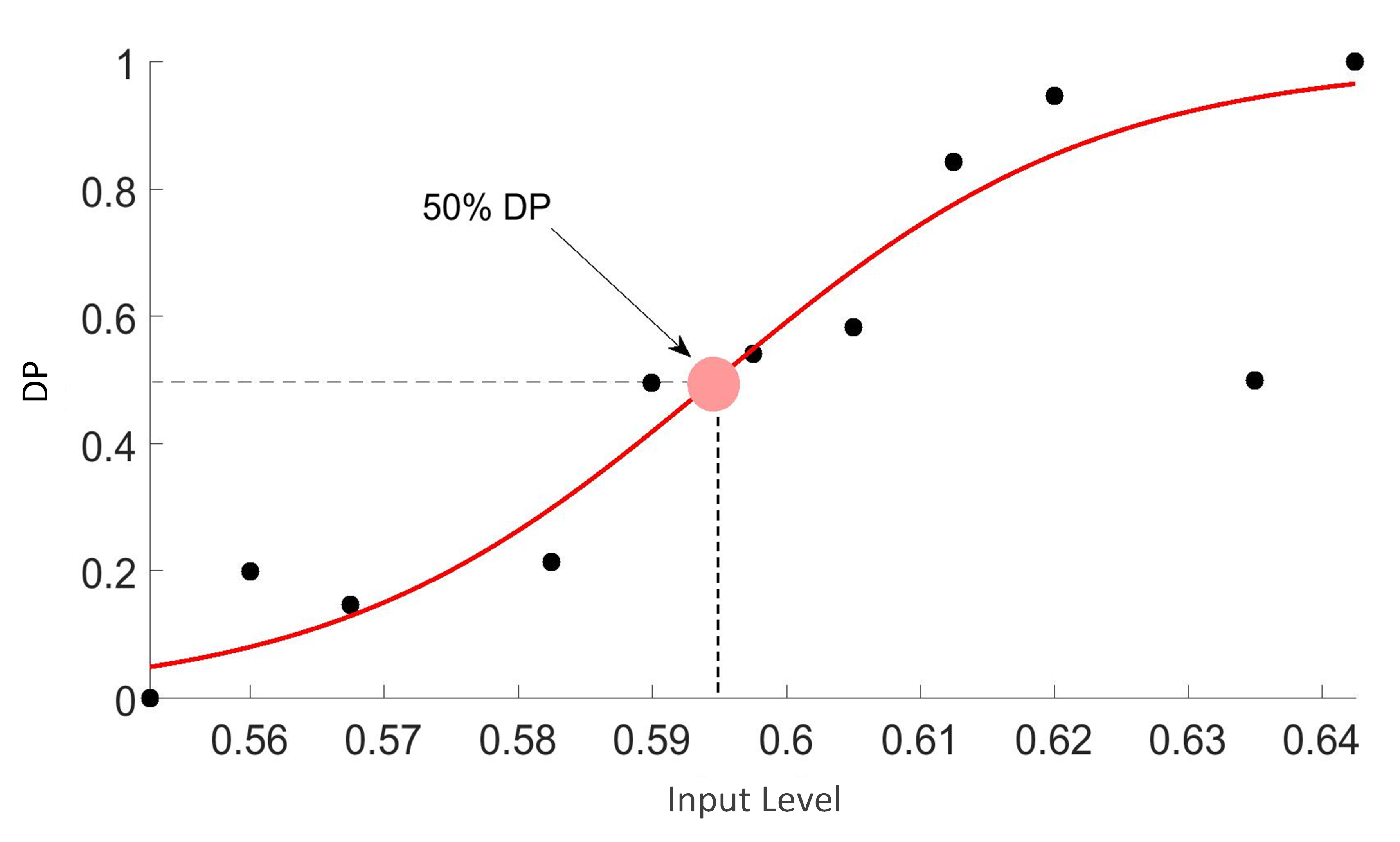}
	\caption[\textbf{Psychometric curve parameters}]{\textbf{Psychometric curve parameters.} The input level at which detection probability is 0.5 defines the threshold \textit{$\theta$}. The steepness of the curve at threshold is represented by the slope \textit{k}, as defined in equation ~\protect\ref{eq:sigmoid}.
		\label{fig:PsycCurveCalc}
	}
\end{figure}

\paragraph{}The Detection Probability (DP) was computed as the fraction of detected trials with input levels in bins of width 0.075  (figure~\vref{fig:PsycCurveCalc}).
A continuous curve of DP as a function of input level was estimated from these points using weighted curve fitting to a sigmoid function: 
%spanning between 0 to 1, represented by the formula:
\begin{align}
	\label{eq:sigmoid}
	DP=f(x;\theta,k)  =
	%\equiv  
	\frac{1}{1+10^{k(x-\theta)}}
\end{align}
The weight of each bin in the fitting was determined by the number of samples in the bin. 
The fitting process was iterative using \textit{Matlab} function for nonlinear fit (nlinfit.m).

\subsection{Model details and parameters}
\label{ModelDetails}

The model is defined by Eq. \ref{eq:Pi} in the main text. Parameters of the basic sigmoid $k,\theta$ were taken from the average of the data. The adaptation term $A$ was modeled as a linear force balancing the recency bias according to past history of the input signal. Specifically, the past history trend $\overline{x}_i$ is defined as
\begin{eqnarray}
\overline{x}_i = x_i(1-e^{-\frac{1}{\tau}})+\overline{x}_{i-1}(e^{-\frac{1}{\tau}})
\end{eqnarray}	
\noindent with $\tau=24$ [time~steps]. The force constant was chosen $\gamma=0.25$. Parameters of the sigmoid are $k=30, \theta=0.595$, corresponding to the average estimated over all human observers. The recency bias parameter is $\delta P = 0.1$. In principle probability was clipped at $[0,1]$ but with the stimuli and response functions used in our experiment this was not required.

%%----------------------------------------------------------------------------------------
%%	RESULTS 
%%----------------------------------------------------------------------------------------

\section{Results}
\label{sec:Results}
\subsection{Sharper psychometric curves for slowly varying inputs}
\label{results:slopes}
\begin{figure}[H]
\centering{
\subfloat{\includegraphics[scale=1,trim = 1mm 0mm 3mm 2mm, clip]{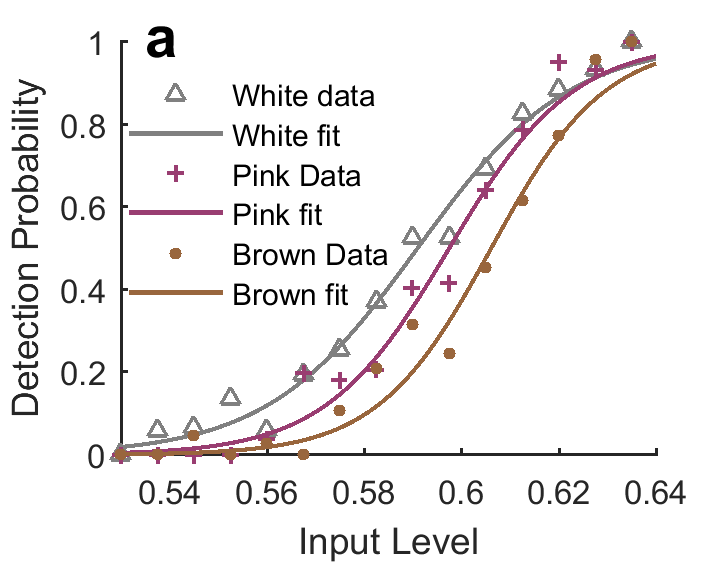}\label{subfig:SlopeExample}}
\subfloat{\includegraphics[scale=1,trim = 1mm 1mm 2mm 2mm, clip]{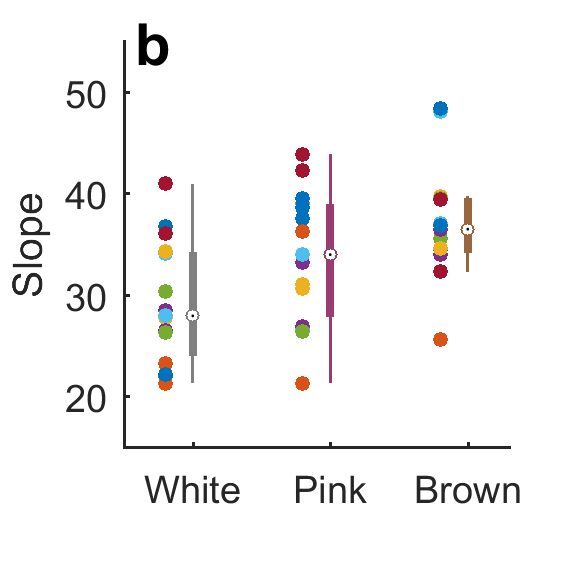}\label{subfig:AllSlopes}}
\subfloat{\includegraphics[scale=1,trim = 2mm 1mm 2mm 2mm, clip]{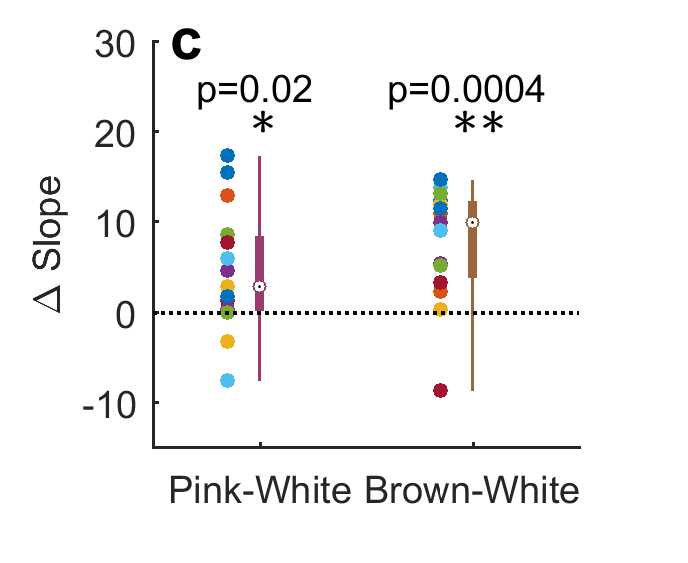}\label{subfig:Slope Relative to White session}}
\caption[\textbf{Slope of psychometric curve depends on input temporal structure}]
{\textbf{Slope of psychometric curve depends on input temporal structure}.
~\protect\subref{subfig:SlopeExample} 
%\textbf{Example of psychometric curve}
Example of psychometric curve for a single observer.  $IR\_F\_23$. 
%Slope is changing between sessions, psychometric curve is steepest in Brown session.
Symbols: binned detection probability, lines: fitted sigmoid. 
Color code marked in legend.  
\protect\subref{subfig:AllSlopes} 
%\textbf{Slope of psychometric curve}
Estimated slopes of psychometric curves for all observers (each colored circle is an individual).
On average, the slope increases for more slowly varying input signals: $29.3\pm2.5\quad32.8\pm4.2\quad36.5\pm2.7~$ for White, Pink and Brown respectively. Errorbars mark the standard deviation (STD) across subjects.
~\protect\subref{subfig:Slope Relative to White session}
%\textbf{Slope Relative to White session}
Individual slopes relative to White session: 
Slope estimated for White signal is subtracted from slopes of the other signals for each observer individually. 
%Differences between sessions within every subject are significant across subjects. 
Statistical T-test performed, significant changes are marked with asterisks with p-values noted.\label{fig:Slope}}
}
\end{figure}

We first characterized the observers' performance to the different temporal signals by estimating the psychometric curve for each of the stimulus types: White, Pink and Brown. 
The psychometric curve is based on the fraction of stimuli detected as a function of input level, and represents the response to the momentary input averaged over the entire experiment. Fig. \protect\ref{subfig:SlopeExample} shows an example of the three psychometric curves computed for one observer.
The curve is most shallow for the White stimulus, where input levels are presented independently at each trial. 
It becomes sharper for the Pink stimulus with temporal correlations, and is sharpest for the Brown stimulus which varies most slowly. 
Sigmoidal fits to the data are shown in solid lines, defining two parameters for comparison among observers: the slope $k$ and the threshold $\theta$.
%\textit{Sl.} and a \textit{Th.}. 
These were extracted for all experiments as described in Statistical Analysis section. 
The slopes for all observers are shown in figure~\protect\ref{subfig:AllSlopes}, where the average is seen to increase with the signal correlation: on average over all observers,
%\begin{align}
\begin{equation}
\label{eq:SlopeRel}
\theta_{White}\leq \theta_{Pink}\leq \theta _{Brown}
%\end{align}
\end{equation}

Since variability between subjects was high, we considered the relative change among stimulus types for each individual separately. For each observer we subtracted the slope of the psychometric curve obtained for the White stimulus from  those of the correlated stimuli, Pink and Brown. 
The result shows with high significance that the per-subject slope of the psychometric curve in response to the White stimulus is lower than that of Pink or the Brown (figure~\protect\ref{subfig:Slope Relative to White session}). 
The slope obtained for the Brown stimulus session was not significantly higher than of the Pink.

The threshold values  $\theta$ of the psychometric curve, in contrast, showed no consistent change among stimuli of different temporal structure (figure~\ref{fig:ThresholdNoChange} in Appendix). 
Consistent with this observation, the total detection probability of individual observers also did not change systematically among the different stimulus regimes.

The significant dependence of psychometric curve slope on temporal stimulus properties provides our first evidence for the sensitivity of perception to these properties. 
Since the different stimuli have the same overall distribution of input levels (see Methods), a strictly static response would result in the same psychometric curve for all of them. 
%This is because characterization by a psychometric curve disregards temporal structure altogether: shuffling the inputs (and corresponding outputs) randomly in time does not change the psychometric curve.
The different curves therefore show that additional variables related to the sequence of presentation affect perception. To reveal these variables we next go beyond the psychometric curve, and analyze responses in temporal context. 

%%%%%%%%%%%%%%%%%%%%%%%%%%%%%%%%%%%%%%%%%%%%%%%%%%%%%%%%
\subsection{Probability of alternating responses is lower than chance in all stimulus regimes}
\label{subsec:POA}
Previous work has shown that psychophysical experiments with uncorrelated signals reveal a "positive recency" effect: the current response tends to be similar to the previous one \cite{Howarth1956,Luce1982,Abrahamyan2016a}. 
A quantity which measures the magnitude of this effect in binary responses is the probability of alternation (POA), defined as the fraction of reversals in a binary string: 
\begin{equation}
POA \equiv \frac{number\ of\ alternations}{number\ of\ trials -1} 
\label{eq:POA}\end{equation}
Where \textit{'Number of alternations'} means the number of changes, between any one type of response and the other. 
For a random, symmetric uncorrelated binary string, POA is expected to be 0.5; lower values correspond to strings with longer streaks.

\begin{figure}[H]
\centering
\includegraphics[scale=0.6]{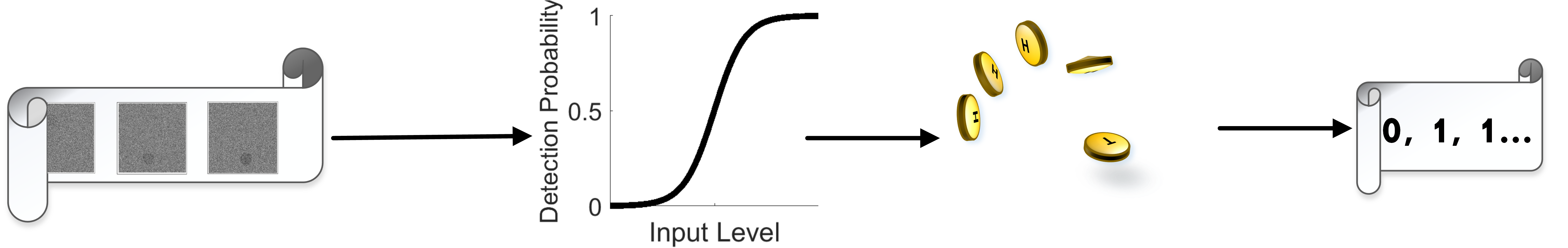}
\caption{\textbf{Instantaneous Model Observers} were simulated to separate effects of the input structure from internal biases. Instantaneous model observers receive a stream of input signals (left) which goes through a local sigmoid input-output relation to define a probability of detection for each input (middle). The response is then determined by a coin-toss according to this probability resulting in a binary detection (right). \label{fig:InstantenousModel}}
\end{figure}

\begin{figure}[H]
	\centering
	\includegraphics[scale=1]{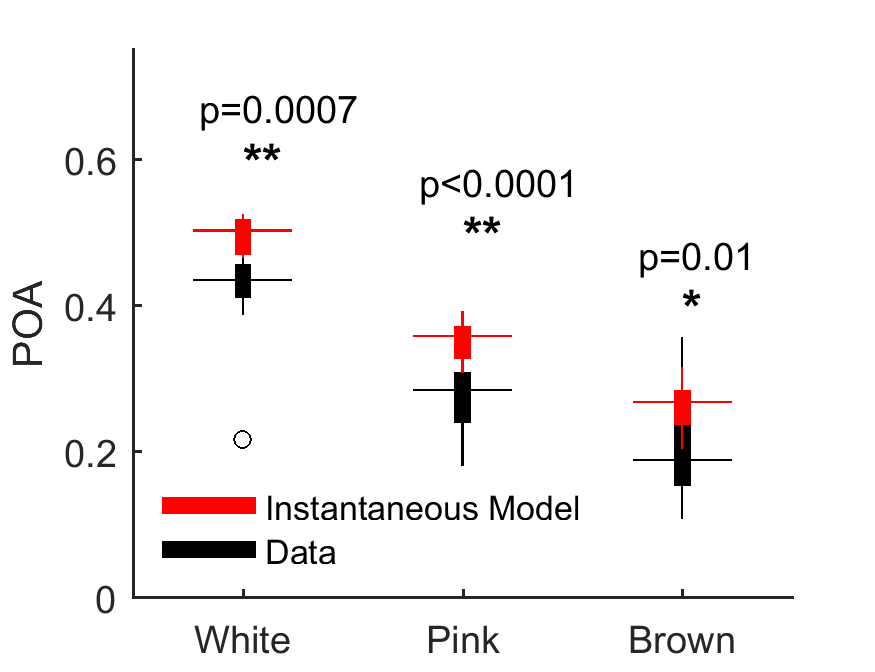}
	\caption[\textbf{Probability of alternations (POA)}]
	{\textbf{Probability of alternations (POA)} 
		computed for human observers (Black) and for 15 instantaneous model observers (Red).
		Means (horizontal lines) are [0.43,0.28,0.21] for the model and [0.47,0.33,0.23] for the data, for White, Pink and Brown inputs respectively.
		Errorbars: 95\% confidence intervals. 
		%Mean and confidence values in the data are for white $0.43\pm 0.02$; 
		%[0.4103 0.4574], 
		%pink 0.28 [0.2400 0.3096] and Brown 0.21 [0.1538 0.2685], and in the model White 0.49 [0.4694 0.5180], Pink 0.35 [0.3272 0.3717]  and Brown 0.26 [0.2365 0.2841].
		Two sided T-test was performed for the difference between the groups of values  (p-values). 
		%The results did not change if the average psychometric curve was used for all stimuli (see appendix, section POA in instantaneous model is not sensitive to variable slopes figure~\ref{fig:POA_sameslope}).
		\label{fig:POA_diffslope}}
	
\end{figure}

In our experiments, the number of alternations is affected also by the input:
if it is correlated in time and varies slowly, it has long stretches of below- or above-threshold regimes, that will result in long streaks of '0` or '1`s, respectively. 
Therefore the POA will be lower for the more slowly varying inputs even for a static observer with no biases, reflecting a property of the input itself. 
To disentangle this input-dependent effect from possible bias and history-dependence of the response, we compare the measured human responses to those of instantaneous model observers. These compute the DP by an 
instantaneous input-output relation and toss a coin with this probability to determine whether the input was detected or not (figure \ref{fig:InstantenousModel}). 
% and thresholds were fixed at the average of all experiments. 
Figure~\protect\ref{fig:POA_diffslope} shows the computed POA for a group of instantaneous model observers, simulated with inputs from each of the three stimulus regimes, marked in red. 
As expected, the white stimulus induces a chance-level POA of approximately 0.5, while the slower stimuli elicit less alternation even without any history-dependence on the observer's part. 
The same figure shows also the POA computed from the experimental data of human observers, marked in black, showing values lower than the instantaneous model for all stimulus regimes.
This reflects an inherent tendency of the observers to repeat the same response as the previous one, generating streaks longer than justified by the input. 
The effect is similar in magnitude across all three stimulus regimes, with experimental POA approximately 0.08 lower than instantaneous model.

%%%%%%%%%%%%%%%%%%%%%%%%%%%%%%%%%%%%%%%%%%%%%%%%%%%%%%%%%%%
\subsection{Different slopes in psychometric curves conditioned on response alternation}
\label{subsec:StayChange}
\begin{figure}[H]
\centering
\subfloat%Conditioned psychometric curevs]
{\includegraphics[scale=0.9,trim =0.75cm 0mm 1cm 0mm, clip]{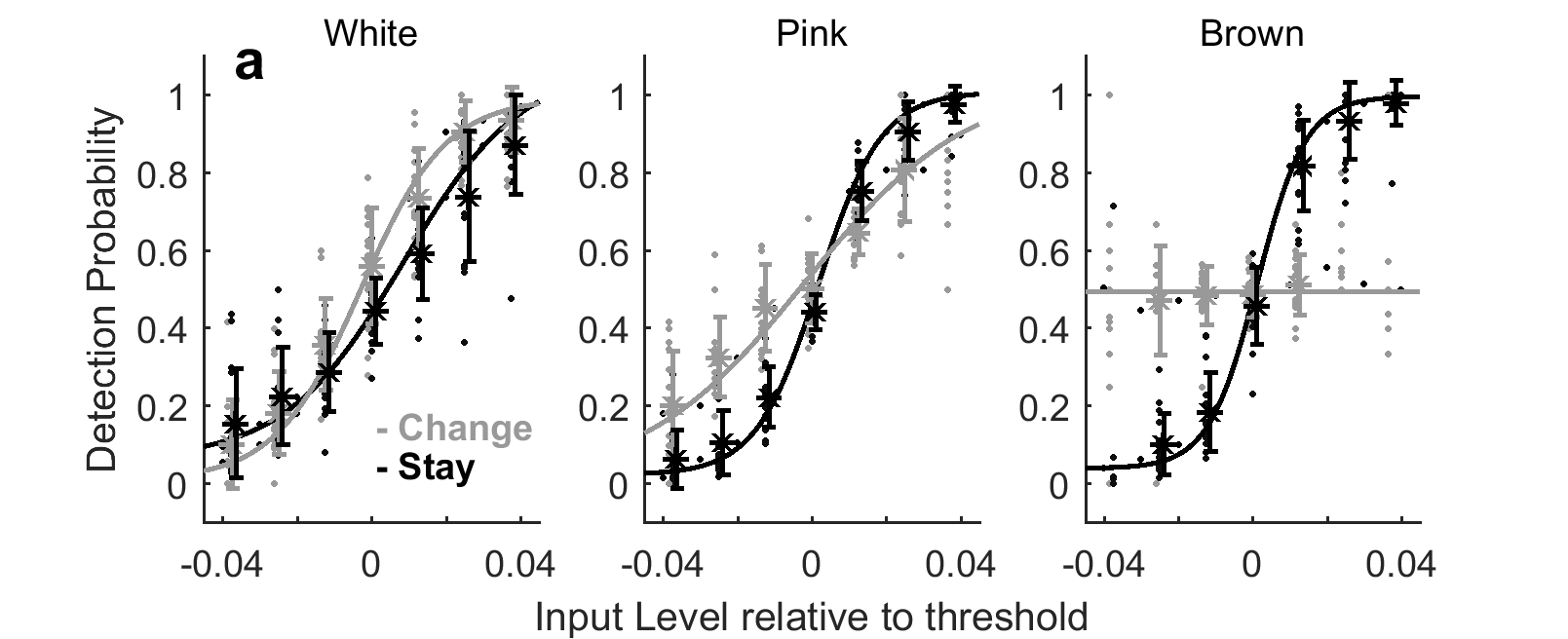}\label{subfig:ConditionedPsychometricCurevs}}
\subfloat%[Subjects fit slopes]
{\includegraphics[scale=1]{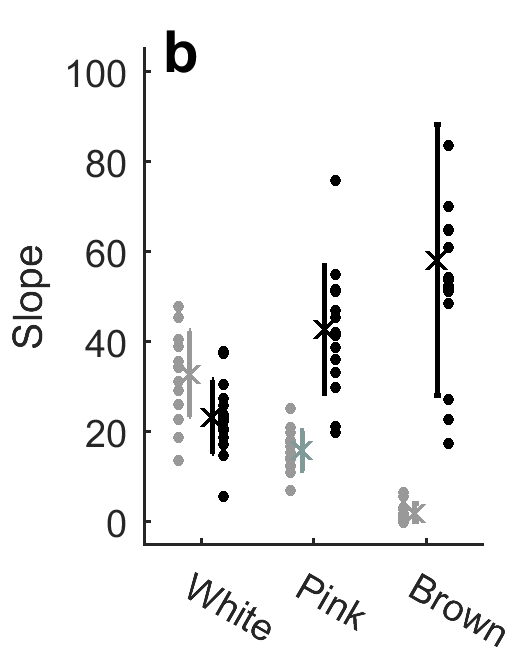}\label{subfig:ConditionedFitSlope}}

\caption[\textbf{Psychometric curves conditioned on response alternation}]{\textbf{Psychometric curves conditioned on response alternation.} 
\newline\protect\subref{subfig:ConditionedPsychometricCurevs}
Trials were divided to two groups: those where the response stayed the same (black) or changed (grey) compared to the previous trial. 
Conditioned psychometric curves have different slopes; the magnitude and direction of the effect is different in the three stimulus regimes. 
\newline\protect\subref{subfig:ConditionedFitSlope} 
Slopes of conditioned psychometric curves fitted to sigmoids for each individual observer separately.   
\label{fig:StayChange}}
	
\end{figure}

The results presented above show that, on average, observers tend to switch their response less often than is required by the input stimulus. 
To further characterize the interplay of this bias with the input, we computed the psychometric curves conditioned on the response-alternation variable. 
We divide all trials to those where the response stayed the same and those where it changed compared to the previous trial. 
Fig. \ref{fig:StayChange} shows the conditional psychometric curves for the three stimulus regimes, black for trials preceded by the same response and grey for those following an alternation. 

For the White stimulus, the two curves differ slightly in their slope. Those trials in which the response stayed the same as the previous one, form a psychometric curve with a smaller slope, indicating a weaker coupling with the input. This is in line with the positive recency bias: responses that stay unchanged possibly indicate a motivation other than the stimulus itself.

The curves for correlated inputs reveal a different result: trials where the response stayed the same as the preceding one make up a sharper relation with the stimulus. This would imply that sometimes a change is made without strict relation to the input, suggesting a bias opposite to the positive recency shown above. This effect becomes even more significant for the Brown input, where response alternations seem to occur with no relation to the input, forming a flat psychometric curve. Although the number of alternations in this experiment is small, the effect is still significant. 

The individual slopes of the conditional psychometric curve for all observers are depicted in Fig. \ref{subfig:ConditionedFitSlope}. It is seen that the slope for the stay-conditioned curve increases systematically with stimulus correlation (black), while the slope for change-conditioned curve decreases (grey). The relation between the slopes changes sign as the correlation increases, showing how different stimulus regime highlight different aspects of the internal biases.
%These results taken together reveal that although these results suggest that both positive and negative biases exist, relative to the instantaneous response dictated by the stimulus; these different biases are exposed in different stimulus regimes.    

%%%%%%%%%%%%%%%%%%%%%%%%%%%%%%%%%%%%%%%%%%%%%%%%%%%%%%%%%%%
\subsection{Hysteresis in psychometric curves conditioned on input trend}
We have seen that conditioning the psychometric curve on output temporal sequences reveals a bias with respect to consecutive responses. 
However, since responses are correlated with inputs, these biases can be caused by input temporal sequences rather than (or in addition to), output sequences. 
Therefore we analyzed also psychometric curved conditioned on properties of the input signal other than the instantaneous value.
Trials were divided into two groups depending on stimulus trend preceding the current one - decreasing or increasing trend. 

\begin{figure}[H]
\centering
\subfloat{\includegraphics[scale=1]{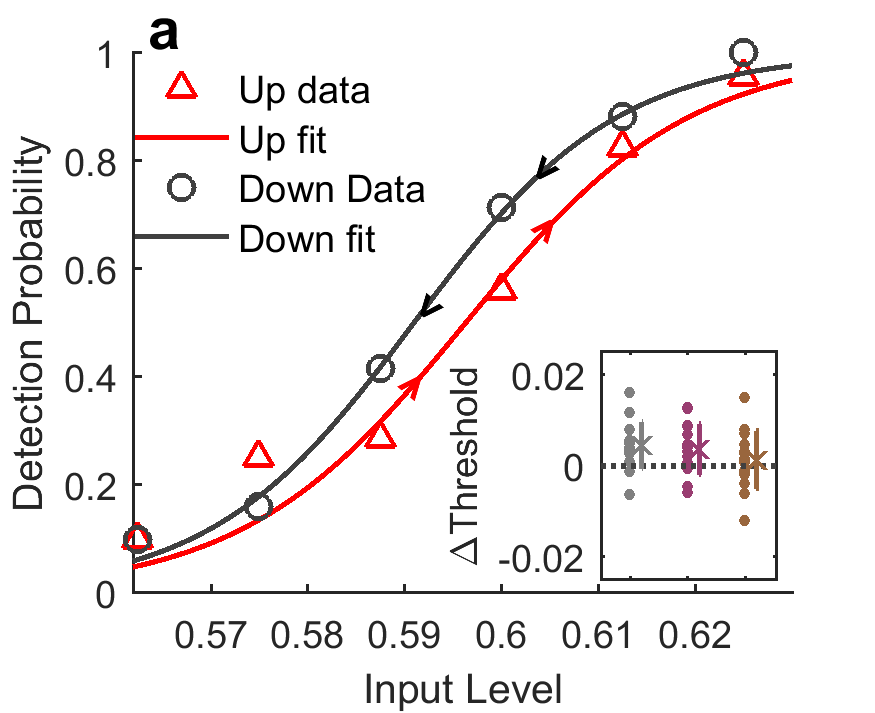}\label{subfig:tau=1}}
\subfloat{\includegraphics[scale=1]{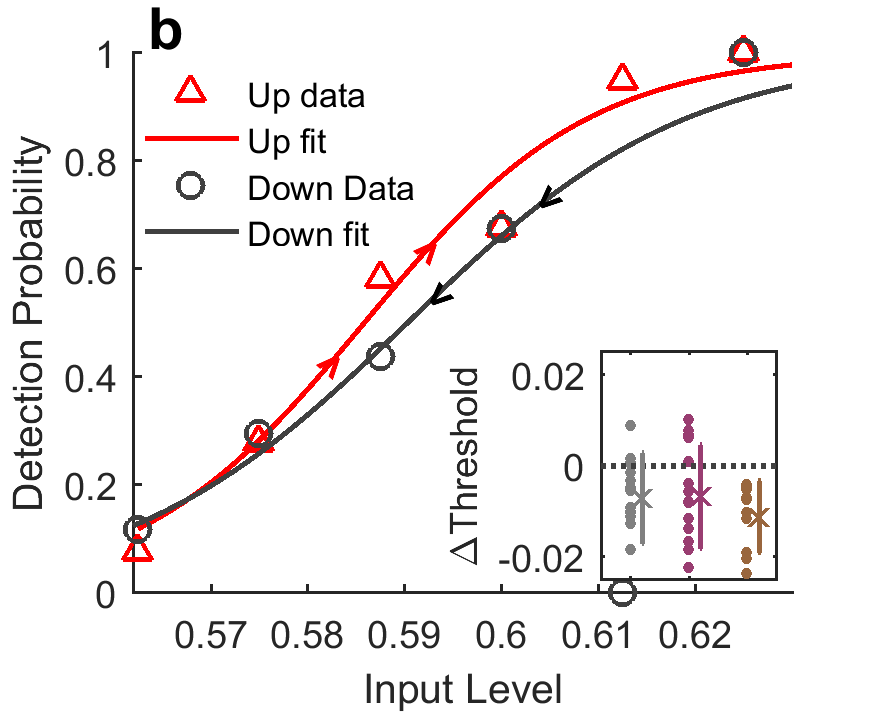}\label{subfig:tau=32}}\\
\centering
\subfloat{\includegraphics[scale=1]{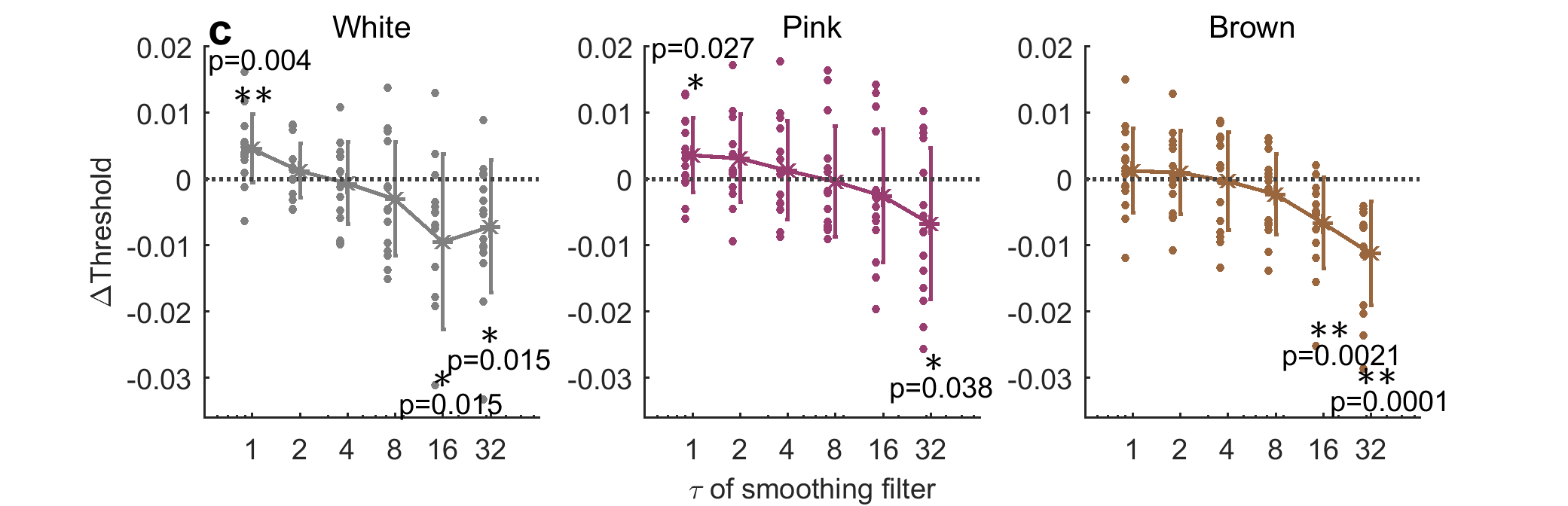}\label{subfig:taus}}
\caption[\textbf{Positive and negative hysteresis}]{\textbf{Positive and negative hysteresis in conditioned psychometric curves.} 
\protect\subref{subfig:tau=1} Example of Up (red) and Down (black) psychometric curves, conditioned on whether the current stimulus is higher or lower than the previous one, in a White stimulus experiment for one observer. 
Data points: circles, sigmoid fits: lines. 
Inset: value of hysteresis, defined as the difference between thresholds, for all observers in the three stimulus regimes (White, Pink, Brown from left to right).  
\protect\subref{subfig:tau=32} %\textbf{Negative hysteresis with $\tau=32$} 
The same analysis as in \protect\subref{subfig:tau=1}, where Up and Down are determined by comparing the current input level to its trend over $\tau=32$ previous trials. Hysteresis is negative here: Up has lower threshold than Down.
%Example data is taken from ``Brown" paradigm with  $\tau=32$. 
Inset: Hysteresis value for all observers in the three stimulus regimes with $\tau=32$.
%\newline Examples~\protect\subref{subfig:tau=1}\&\protect\subref{subfig:tau=32} are taken from subject IR\_F\_23.
\protect\subref{subfig:taus} 
%\textbf{Hysteresis with various $\tau$ values} 
Hysteresis value of all observers in the three stimulus regimes, as a function of the timescale defining the past trend ($\tau$). 
T-test performed against a null hypothesis of zero hysteresis; significant values are marked with asterisks and p-values noted.
%\textcolor{red} {asterisks are not visible. They are swamped in the data points. maybe change color?}
}\label{fig:Hysteresis}
\end{figure}

Figure~\ref{subfig:tau=1} shows the results for a white input signal, where the two groups correspond to the current stimulus being either larger or smaller than the previous one. 
Red circles, "Up data", show trials in which the current stimulus was higher than the previous one. 
Black circles, "Down data", is constructed from trials in which the current stimulus was smaller. Solid curves show sigmoidal fits to the two data sets.
This analysis reveals a positive hysteresis: the Up curve has a higher threshold than the Down curve. 
Such positive hysteresis indirectly reflects a tendency to perceive the input as similar to the previous one: for the same input level, coming from high stimulus our perception is higher than coming from a previously lower one. 
The difference between the thresholds of the two conditional curves is shown in the inset, for all three stimulus regimes. 
It is seen that the effect is strongest for White input and decreases to an insignificant value for the Brown input. 
%This can be partially explained by the fact that, in a slowly-varying stimulus, changes between consecutive input levels tends to be very small, whereas in the white stimulus they can be of any magnitude.

This analysis reveals a sensitivity to the change in stimulus, but takes into account only the current and previous trials.
A generalization which takes into account a longer history, is achieved by comparing the current stimulus to the past history over a timescale $\tau$. 
Trials are then divided into two groups depending on whether the current input value is higher or lower relative to the general past trend. 
Specifically, the input levels are filtered using an exponential filter of time constant $\tau$; the result is then compared to the current input level. 
We note that time is here measured in number of trials. 
%\begin{align}
%\label{eq:filter}
%F(x_n)=x_n  {(1-e^{-\frac{1}{\tau}})} + F(x_{n-1})(e^{-\frac{1}{\tau}})
%\end{align}

%Each component $n$ of the filtered signal was classified according to the derivative sign $\delta F(x_n)$: labeled ``up" if derivative of filtered signal was positive (level increased) and ``down" if it was negative (level decreased). 

%%\label{eq:updown}
%\begin{subequations}\label{eq:updown}
%\centering{
%\begin{flalign}
%\delta F(x_n)=F(x_{n})-F(x_{n-1}) \\
%n \in\left\{ 
%\begin{array}{lrrr}
%up ~\qquad \,if\quad  \delta F(x_n) \geq 0\\
%%n \in
%down \quad if\quad  \delta F(x_n)<0
%\end{array} \right.
%\end{flalign}
%}\end{subequations}

%Parameter $\tau$ in equation~\ref{eq:filter} determined how far back the history of inputs counted in defining the trend;

%Simply ``up" when $x_n\geq x_{n-1}$ and ``down" if $x_n<x_{n-1}$.
%\newline As the the filter was getting longer, with higher $\tau$ values, the ``up"/``down" relied to more general trend of the input signal. 

Psychometric curves were estimated independently for the two groups, Up and Down, as before. 
An example is shown in figure~\ref{subfig:tau=32}, where a timescale of $\tau=32$ was used for defining the past input trend in a White stimulus experiment.
In contrast to the previous plot, here we find a negative hysteresis effect, namely the threshold for Up trials is lower than for Down trials. 
A similar effect was found for all stimulus regimes. 
Negative hysteresis reflects an increased sensitivity moving from a weak to a stronger stimulus, which is usually referred as adaptation. 
Such adaptation is typical to slowly changing environments that allow reliable prediction, with deviations from that prediction resulting in an enlarged reaction. 
%Indeed, this effect was prominent in Brown session where input levels were actually changing slowly. 

Quantifying the degree of hysteresis as the difference between thresholds of Up and Down curves, allows us to plot this difference for a range of $\tau$ values, corresponding to the length of history defining the trend. 
Figure. \ref{subfig:taus} shows the result of this analysis, depicting all individual observers as dots, with averages and standard deviations marked. 
The trends are clear and similar for all stimulus regimes: in the short term a positive hysteresis appears, which decreases with $\tau$ until it eventually crosses over to a negative hysteresis for long times. 
The White stimulus reveals the largest magnitude of positive hysteresis, whereas for the Brown stimulus the negative hysteresis dominates. 
These results show that both processes, positive and negative biases, exist in human observers. 
It appears that they emerge with different characteristic timescale - positive bias over short times and negative bias over longer times. 
The ultimate response pattern results from an interplay of the two and the nature of the stimulus.
%\begin{align}
%\label{eq:HystDef}
%Hystersis=Th._{up}-Th._{down}
%\end{align}

%%---------------------------------------------------------------
%%	Model
%%---------------------------------------------------------------
\subsection{A Model for biased perception}
\label{sec:model}
%\subsection{General} 
The experiments presented above suggest that, in addition to the input signal, two inherent opposing forces act to shape perception  on different timescales. 
On one hand, human observers tend to stick to their previous responses even when stimuli change. 
On the other hand, over longer timescales, an adaptation effect occurs which effectively keep the observer from sticking to a constant response for too long. Using the model of fig.~\ref{fig:InstantenousModel} as a basis, one may describe these two effects as modulations of the instantaneous input-output function. The positive recency can be modeled by adding a small probability bias $\delta P$ to the output, positive if the previous input was detected and negative if it was not. Adaptation can be described by a modification of the sigmoid threshold, or equivalently, by adding a bias to the perceived signal~\cite{BendaHerz2003}. Together these two biases give the following probability $P_i$ of detecting the input signal at trial $i$:
%The incoming signals go through the sigmoid response function $f(z)$ defined by 
%\begin{eqnarray}
%f(z)=\frac{1}{1+10^{k(z-\theta)}},
%\end{eqnarray}
%which in the simple case also determines the detection probability $P_i$. 
\begin{eqnarray}
P_i=f(x_i-A)+\delta P, \quad\quad A=\gamma(\overline{x}_i -\theta)\label{eq:Pi}
\end{eqnarray}
where $\overline{x}_i$ is the input signal filtered over some timescale $\tau$ into the past (see Methods for details). This adaptation effectively adjusts the threshold as a result of transiently high or low inputs. 

%Therefore, adaptation is applied in the sensory stage;
%The input signal is \textit{linearly} filtered before it passes through the nonlinear (sigmoid) function, this is known in the signal processing context as a "linear-nonlinear`` model~\cite{Narendra1966}.
\begin{figure}[H]
	\centering
	\includegraphics [scale=0.6,trim = 0.5cm 0cm 0cm 0.2cm, clip]{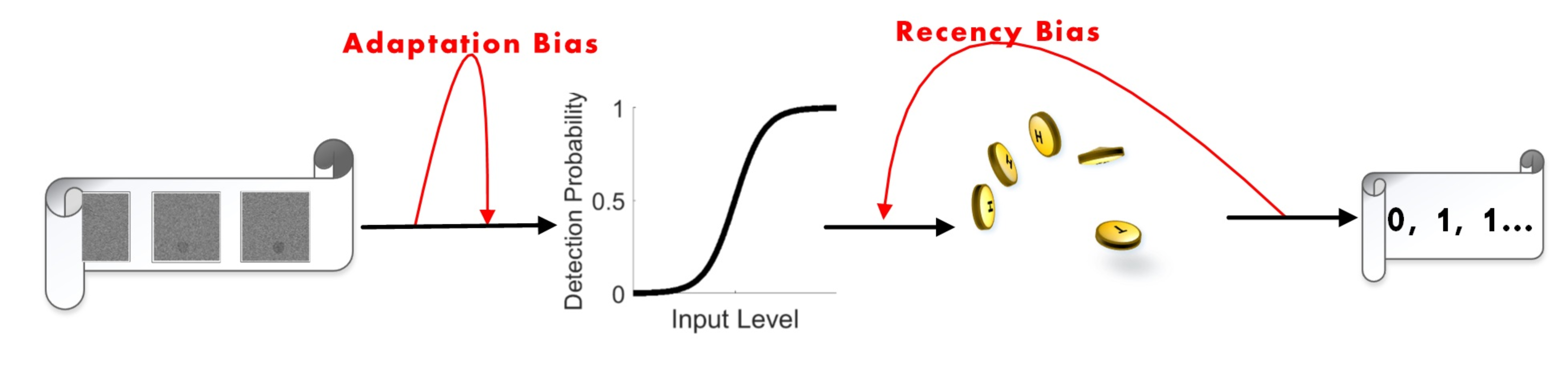}
	\caption[\textbf{Sensory-Cognitive model - dependent on history of inputs and responses}]
	{\textbf{Sensory-Cognitive model.}
		The backbone structure of the model (black arrows) is composed of a fixed input-output relation (\textit{sensory process}) determining the probability of detection, and a coin flip decision based on this probability \textit{cognitive process}. 
		Two biases modulate this backbone (red arrows): an \textit{Adaptation Bias} varies the threshold based on the input history; and a \textit{Recency Bias} modifies the final decision based on previous responses.
	}

	\label{fig:SensoryCognitiveModel1}
\end{figure}

\paragraph{}

The general structure of the model is depicted as a black backbone in Fig. \ref{fig:SensoryCognitiveModel1}, with the history-dependent modifications added as red arrows. 
Here adaptation acts directly on the input stream whereas the cognitive process is affected only by the previous output. 
This partition of the two history-dependent modifications is consistent with recent fMRI experiments \cite{Schwiedrzik2014}, indicating that they are mapped to distinct brain regions: 
Adaptation was linked to primary visual areas whereas positive recency was to high cognitive areas.

\begin{figure}[H]
	\centering{
		\subfloat{\includegraphics[scale=1]{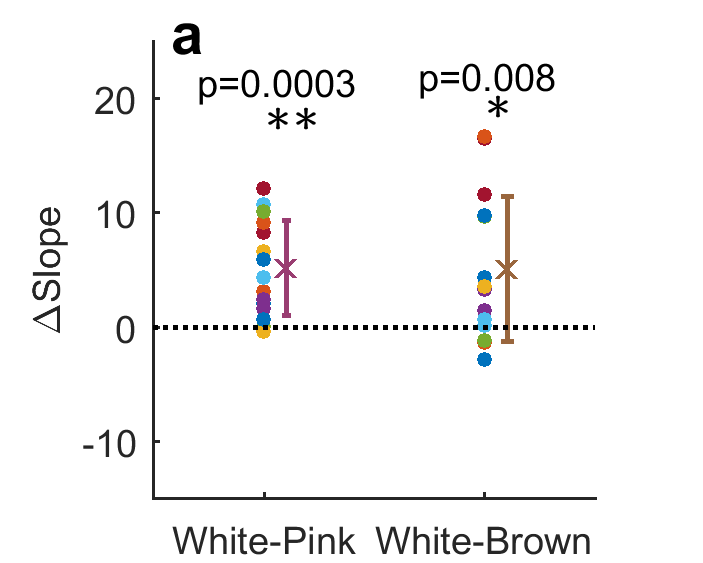}\label{subfig:Modelslopedif}} % [Slope of Model subjects]
		\subfloat{\includegraphics[scale=1]{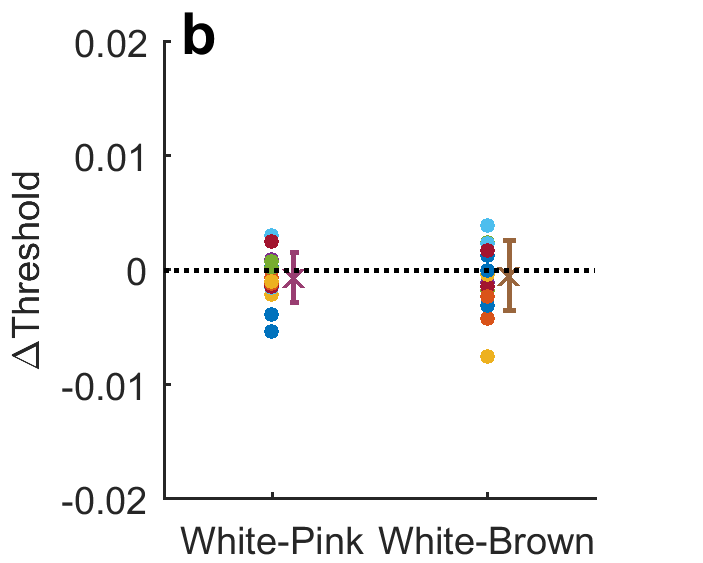}\label{subfig:Modelthresholdif}}%[Threshold of Model subjects]
		\caption[\textbf{Psychometric curve parameters in the model}]
		{\textbf{Empirical psychometric curve parameters for model observers.} 15 model observers were presented with the three stimulus types, and their responses analyzed to estimate empirical psychometric curves. Relative slopes %(~\protect\subref{subfig:Modelslopedif}) 
			\protect\subref{subfig:Modelslopedif} and thresholds \protect\subref{subfig:Modelthresholdif}
			%~\protect\ref{subfig:Slope Relative to White session}) 
			were computed by subtracting the corresponding parameters of the White stimulus from those of the Pink and Brown. Compare to experimental results in Fig. \ref{fig:Slope}.
			%~\protect\subref{subfig:Modelthresholdif}. 
			%Individual threshold with no change across all sessions (similar to human subjects in figure~\protect\ref{subfig:ThresholdR2W}).}
		%~\protect\ref{fig:ThresholdNoChange}).
	}
		\label{fig:Modelslope}}
\end{figure}

\begin{figure}[H]\centering{
		\includegraphics[scale=1]{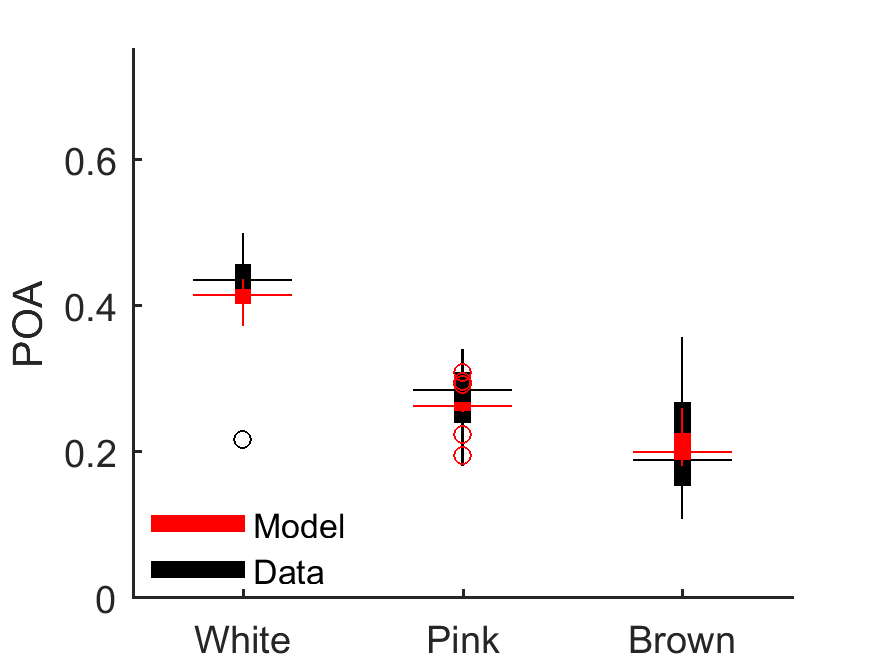}
		\caption[\textbf{POA of model and data}]{\textbf{POA of model and data} Probability of alternation (POA) in the response of humans subjects (Black) and of model subjects (Red).}
		\label{subfig:StreakinessModel}}
\end{figure}

\begin{figure}[H]
	\centering{
		\includegraphics[scale=1]{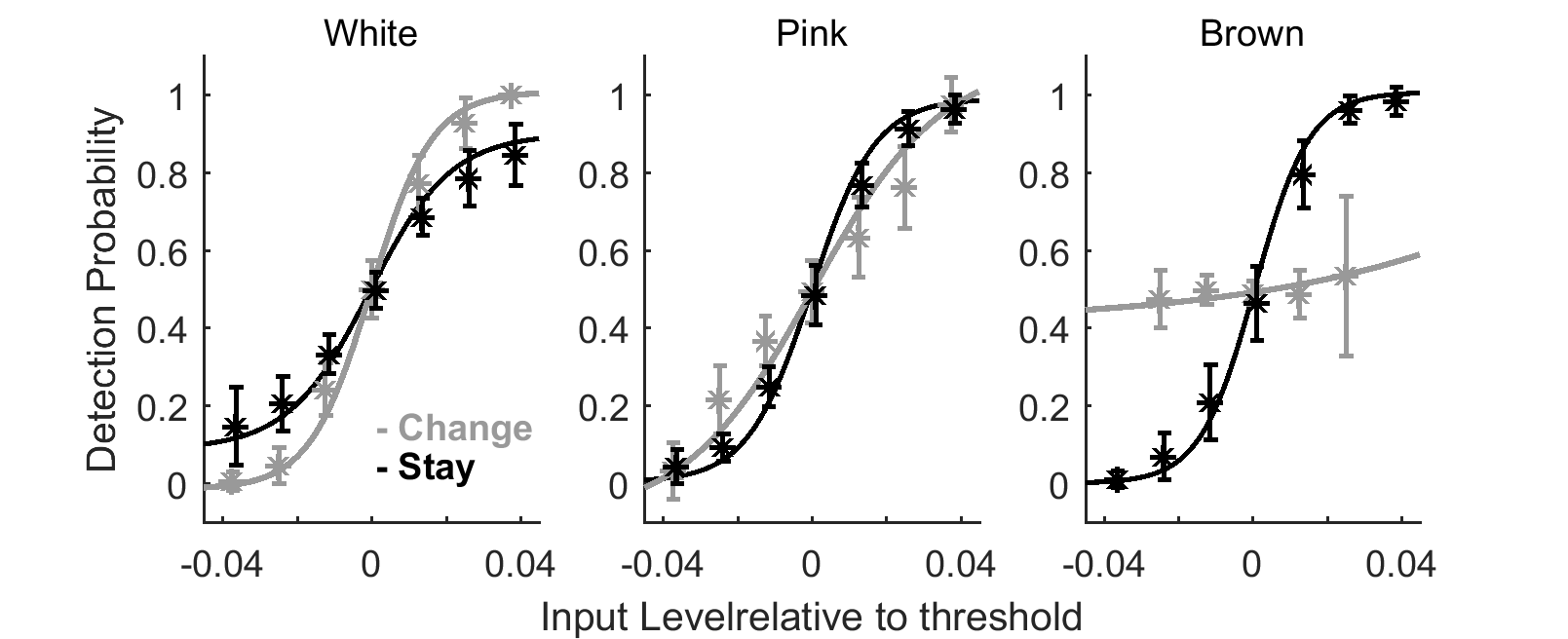}
		\caption[\textbf{Model psychometric curves conditioned on response}]{\textbf{Model psychometric curves conditioned on response alternation.}
Trials were divided into those in which the response was the same as in the previous trial ("stay"; black) and those in which it changed ("change"; gray). Psychometric curves were estimated for each group separately.}
\label{subfig:ModelStayChange}}
\end{figure}

\begin{figure}[H]
	\centering{
		\includegraphics[scale=1]{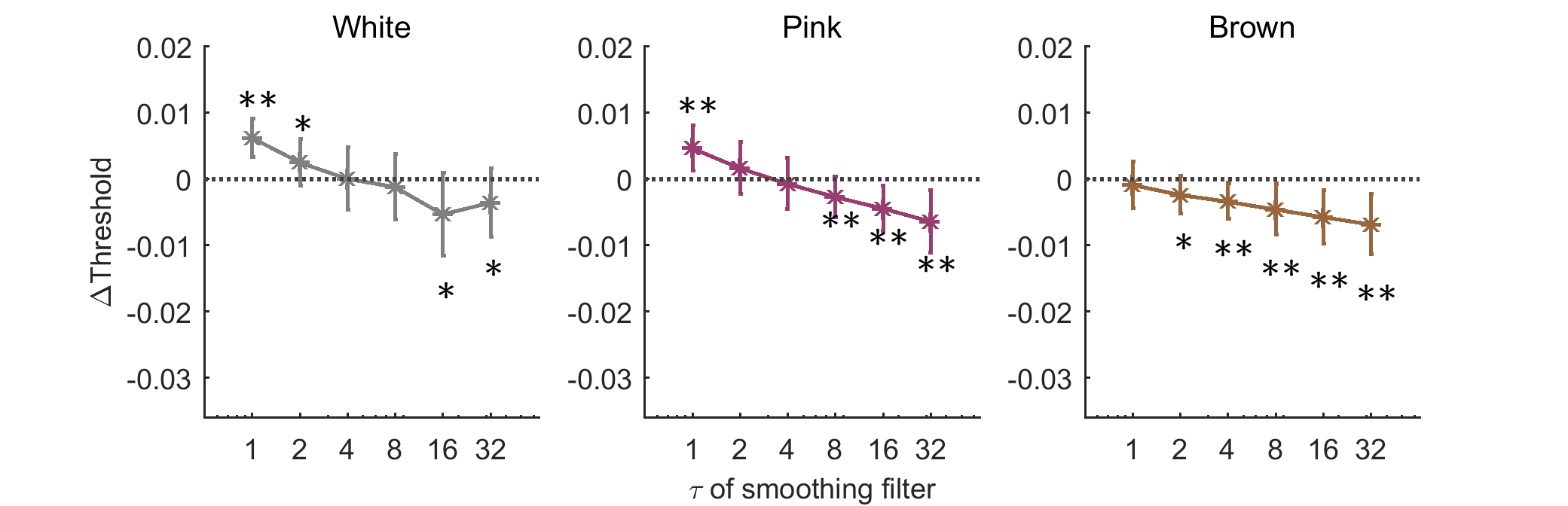}
		\caption[\textbf{Hysteresis in model}]{\textbf{Hysteresis in model psychometric curves conditioned on stimulus trend} Hysteresis in model shows qualitatively the same behavior as a function of $\tau$, the timescale used to define the stimulus trend as Up or Down, as human observers, for all stimulus regimes}
		\label{fig:ModelHysteresisintaus}}
\end{figure}

We used this model to generate a set of 15 observers, and simulated the same protocol as the experiment. 
Input signals were synthesized in the three stimulus regimes, presented to the model observers, their responses (0/1) recorded, and the sequences of input and output were analyzed similar to the experiments. 
The results are presented in Figs. \ref{fig:Modelslope}-\ref{fig:ModelHysteresisintaus}.

First we used the responses of the model observers to construct their empirical psychometric curves. 
Although the input-output sigmoid function defined in the model (middle box in Fig. \ref{fig:SensoryCognitiveModel1}) was fixed, the existence of history-dependence in the model together with different temporal structure of the inputs resulted in empirical psychometric curves that depended on the stimulus regime. 
Fig. \ref{fig:Modelslope} depicts the parameters of these empirical curves for the two correlated inputs - Pink and Brown - after subtracting those computed for the White stimulus. 
The model produced sharper psychometric curves (larger slopes) for the slowly varying stimuli (\protect\ref{subfig:Modelslopedif}), while the threshold values remain unchanged (\protect\ref{subfig:Modelthresholdif}). 
The effects are similar in direction and in magnitude to that found for human observers. 

Next we considered the probability of response alternation (POA) averaged over the entire experiment, for the different stimulus regimes. Fig. \ref{subfig:StreakinessModel} shows the results for all model observers (red) together with the same quantities computed for the experiments on human observers (black).
They are practically indistinguishable, showing that the model captures correctly the tendency for positive recency equally well in all stimulus regimes.

%\subsubsection{Probability Conditioned on Response}
The model also captured the empirical psychometric curves conditioned on the response, as well as those conditioned on the stimulus trends. 
Fig. \ref{subfig:ModelStayChange} shows the results for conditioning on stay / change of the response relative to the previous one (compare to experiment in Fig.~\ref{fig:StayChange}). 
Fig.~\ref{fig:ModelHysteresisintaus} shows the hysteresis - difference in thresholds between the two conditioned groups - for conditioning on the direction of stimulus changed, defined over various timescales. 
This can be compared to Fig. \ref{fig:Hysteresis}: the model shows the same general profile of hysteresis values as a function of the timescale used to define the trend in the stimulus. 

We note that in all model simulations the 15 observers had the same parameters, and therefore variability between them is smaller than between human observers.

Other models that differ in details of the history-dependent effects were also tested. For example, the dependence of the adaptation bias (threshold modification of the input-output curve based on a history of length $\tau$) could be made to depend on the output rather than the input. A sketch of this model is presented in Fig.~\ref{fig:SensoryCognitiveModel2} (Appendix). The results of model observers following these processes were indistinguishable from the one presented above. Other possible combinations were also tested, for example both biases modulating the cognitive processes (Appendix); however, this was found to be provide a slightly worse fit to the data, and moreover was less robust, i.e. more sensitive to the choice of parameters.

%They used a modified version of a paradigm introduced by Gepshtein and Kubovy (2005)\cite{Gepshtein2005} which allows the manipulation of both phenomena concurrently.
%It is to be noted that adaptation as presented in their experiment is not exactly the same as ours since they relay only to the effect of one last stimulus.
%\subsection{Sensory-Cognitive model - dependent on history of inputs and responses}\label{subsec:SensoryCognitiveModel1}

%\begin{figure}[H]
%\centering{
%\includegraphics[scale=0.18]{./graphics/image002.jpg}
%\caption[\textbf{Effect of 2 projection planes}]{\textbf{Effect of 2 projection planes} On the left (Blue) the effect of instantaneous level when input contains history.
%On the right (Green) effect of $1^{st}$ derivative of the input, which is also modified by the input temporal structure.}
%\label{fig:2planes}}
%\end{figure}

%--------------------------------------------------------------
\section{Discussion}

In this study we have used temporally structured stimuli to investigate dynamic aspects of fluctuations in visual perception. The task was a visual detection near threshold; observers were asked to report detection / no detection, and no feedback was provided. While this elementary task is simple in itself, the sequence of input levels (contrast level for detection of a circle on a background), varied in time in a nontrivial way, spanning three stimulus regimes: White - where consecutive trials were independent; Pink - where they were positively correlated in time; and Brown - the most slowly varying succession of inputs. 

Characterizing the detection experiments first by a static psychometric curve, we found that responses to more slowly-varying stimuli exhibit a sharper curve (Fig.~\ref{fig:Slope}). This may seem surprising at first, since in all cases the distribution of stimuli was the same, and moreover the total detection probability of observers was also the same.  
However, this result clearly establishes that a static response curve does not tell the whole story of how perception is formed. Indeed, previous work had already found that perception depends on 
sequences of events. The different psychometric curves highlight the fact that these history-dependencies interplay with the input temporal structure in different ways. 

To shed light on this interplay, we analyzed the data using methods that emphasize history dependence. For example, dividing the experiments to groups conditioned on some sequence of events, and computing average responses or full psychometric curves for each of them separately. 
Using this approach, we found two opposing biases affecting perception on two separate timescales. First, a positive recency effect: human observers tend to repeat their previous judgment of the stimulus beyond chance, regardless of the input itself \cite{Verplanck1952,Snyder2015}. This bias was manifested across all stimulus regimes by a decreased probability of alternating response over the entire experiment, when compared to a static observer (Fig.~\ref{fig:POA_diffslope}). It was also apparent indirectly from the hysteresis in psychometric curves conditioned on the direction of recent change of input (Fig.~\ref{subfig:tau=1} and~\ref{subfig:taus}, short timescales); since input is correlated with response, such hysteresis suggests a stickiness to the previous response. This analysis revealed that the 
positive recency effect is strongest for the White stimulus regime, and decreases as input correlations increase. It also showed that positive recency is a relatively short-term effect, i.e. depends most strongly on the previous input. 

From a functional point of view, this bias can be understood as stabilizing perception against fluctuations, by holding a prior representation of regularity in the sensory world. Therefore the White stimulus, which is far from realistic and does not exhibit these regularities, exposes this bias most strongly. In the face of white noise, a tendency to repeat the same response is detrimental to perception; this can be clearly seen by conditioning on the events "stay" or "change" in the response, showing that for White input staying induces a more noisy relation with the input (Fig.~\ref{fig:StayChange}). 

A second and opposite tendency, to change the response, was also revealed in the same set of experiments. This was mostly apparent when input stimulus changed very slowly, namely using the Brown signal. Here a psychometric curve conditioned on the response alternation showed an opposite effect to that of the White input: responses that alternated were so noisy they were practically uncorrelated with the stimulus (Fig.~\ref{subfig:ConditionedFitSlope}). This suggests an inherent tendency to change response after staying "stuck" for a long time. The same effect was also manifested in the response curves conditioned on the input trends over long times, resulting in negative hysteresis (Fig.~\ref{subfig:tau=32}). This can be understood as follows: if the current input was low relative to the recent long-term average, in a slowly varying input, this would imply that the input has been high for a while. Therefore, the shift of the conditional curve to higher values implies an adaptation of the threshold to the recent statistics, moving it to the center of the incoming signal. In signal processing this can be rationalized as an optimal utilization of limited dynamic range of response to best code the incoming signal \cite{Laughlin1981,Brenner2000}. 
In contrast, in binary detection, such a mechanism implies constantly being around the detection threshold, where ambiguity is maximal. Therefore here it can be understood as an exploratory force, keeping the response from falling into a continually repeated pattern. Together, the two internal opposing processes affecting perception allow a flexible balance between positive and negative biases. 

A simple model was developed to describe how these two biases can modify an instantaneous input-output function dynamically depending on history. All experimental results are satisfactorily reproduced by one set of parameters (which did not require very fine tuning). The two opposing processes were modeled as affecting two separate stages of detection - sensory, where the signal is interpreted, and cognitive, where a decision is made. This is in line with previous fMRI experiments indicating these biases are formed in different brain regions \cite{Schwiedrzik2014}. Our analysis revealed further that the two processes are characterized by different timescales. However, exactly what they depend on in terms of history - input or output - did not make a significant difference in the results. This is understood since input and output are expected to be correlated with one another. As a result, two model versions describe the data equally well.

In a more general context our study offers a novel methodology which my be useful in future studies. In terms of experiment design,
the input signal is regarded as a continuous stream and utilizes structured stimuli to create history-dependence. This approach allows dependence on history to surface on several timescales that were not inserted "by hand" into the experiment. On the complementary side of data analysis, we have used conditioning on various sequences of events with different characteristic timescales to expose the relations between temporal input and temporal output. These methodologies can be easily implemented in other sensory modalities and tasks.

%%----------------------------------------------------------------------------------------
%%	BIBLIOGRAPHY
%%----------------------------------------------------------------------------------------
\renewcommand{\refname}{\spacedlowsmallcaps{References}} % For modifying the bibliography 
\bibliographystyle{apalike}
\bibliography{Articlebib1} % The file containing the bibliography

%%----------------------------------------------------------------------------------------
%        APPENDIX
%%----------------------------------------------------------------------------------------
\newpage
\section{APPENDIX}
\label{sec:APPENDIX}

\subsection{Paradigm validation - the task involves no learning}
\begin{figure}[H]
	\centering{
		\subfloat{\includegraphics[scale=1]{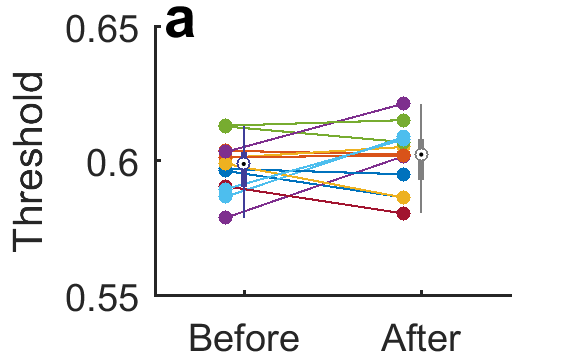}\label{BAThreshold}}
		\subfloat{\includegraphics[scale=1]{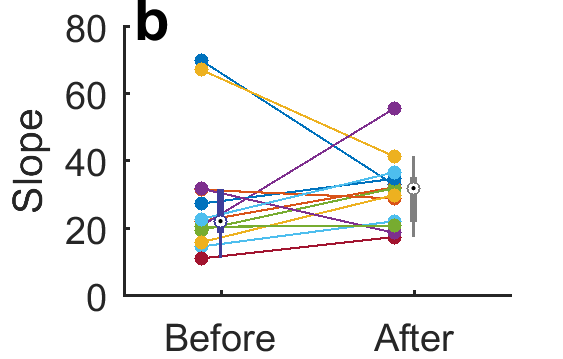}\label{BASlope}}
		\subfloat{\includegraphics[scale=1]{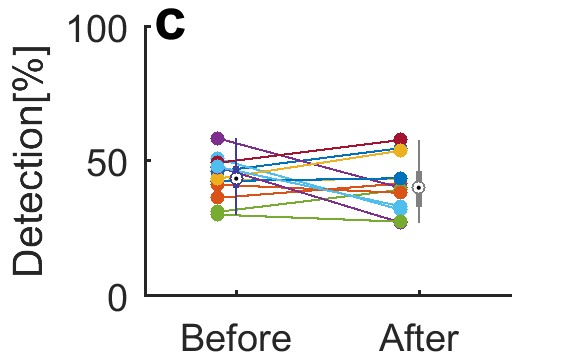}\label{BADP}}
		\caption[\textbf{Performance comparison before \& after experiment}]
		{\textbf{Performance comparison before \& after experiment.} 
			the parameters that were compared are~\protect\subref{BAThreshold} Threshold,~\protect\subref{BASlope} Slope and~\protect\subref{BADP} Total detection rate.
			Each colored circle represents a different subject, errorbars marking mean and standard deviation across subjects. 
			No significant change in any parameter.}
		\label{fig:Before & after}}
\end{figure}

\paragraph{}Control sessions before and after the main experimental sessions were used to verify that no learning was involved. 
Three parameters were tested: the two parameters of the psychometric curve, the threshold $\theta$ and the width $k$, ande the total detection fraction over the session. 
As seen in figure~\ref{fig:Before & after} there is no significant change in any of these parameters before and after the experiment.

\subsection{No change in psychometric thresholds among stimulus types}
\label{subsec:NoChangeTh}
Comparing thresholds of psychometric curves fit to individual observers, we see no change in threshold throughout the experiment. This is true also on average over all observers.
\begin{figure}[H]
\centering
\subfloat{\includegraphics[scale=1]{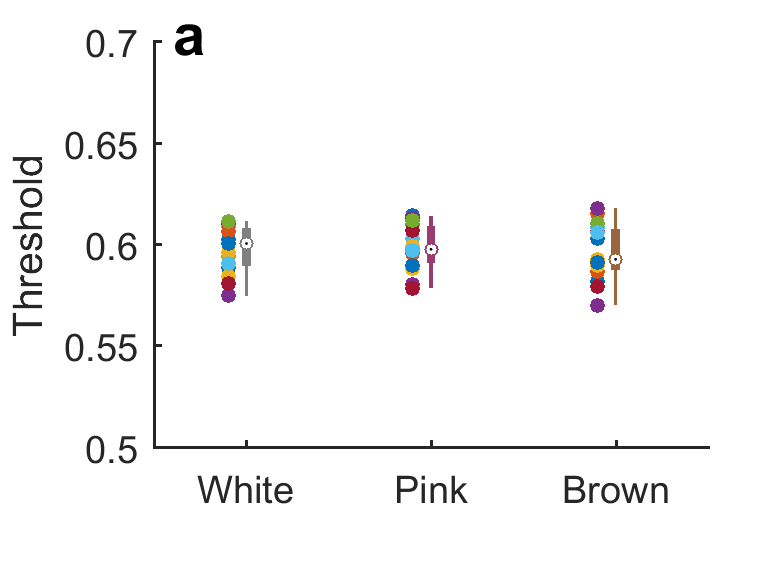}\label{subfig:ThresholdAll}}
\subfloat{\includegraphics[scale=1]{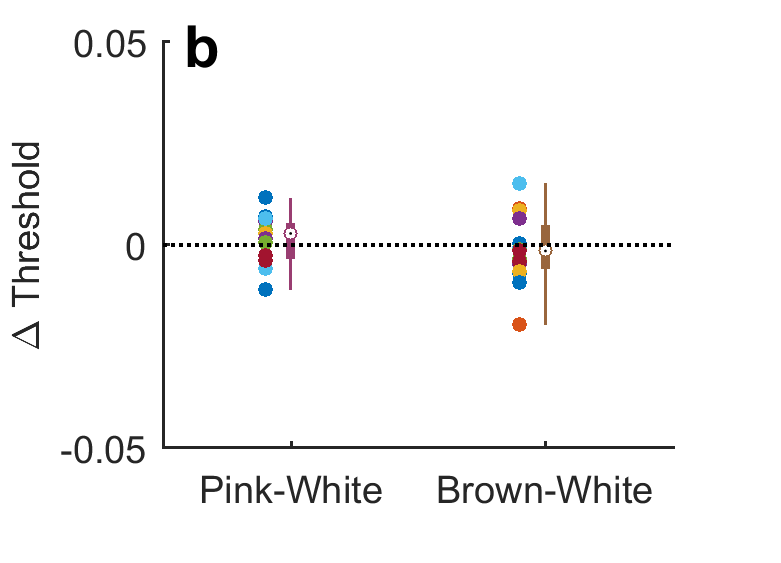}\label{subfig:ThresholdR2W}}
\caption[\textbf{No change in threshold}]{\textbf{No change in threshold between differently correlated input signals.}~\protect\subref{subfig:ThresholdAll} Thresholds of psychometric curves for White, Pink and Brown stimuli. 
~\protect\subref{subfig:ThresholdR2W} Relative threshold obtained by subtracting the threshold of the White stimulus from Pink and from Brown for each observer.}
\label{fig:ThresholdNoChange}
\end{figure}

\subsection{POA in instantaneous model is not sensitive to variable slopes}
\label{subsec:POAnotSensitive2slope}
Comparison of POA between human and model observers, similar to main text results.
Here we used for the instantaneous model psychometric curve the average of the entire data set, i.e. the same curve for the three input signal types ($k =30$).
The results reported in the main text results  that refer to slopes  are insensitive to this modification of the model instantaneous observer.
\begin{figure}[H]
\centering
\includegraphics[scale=1]{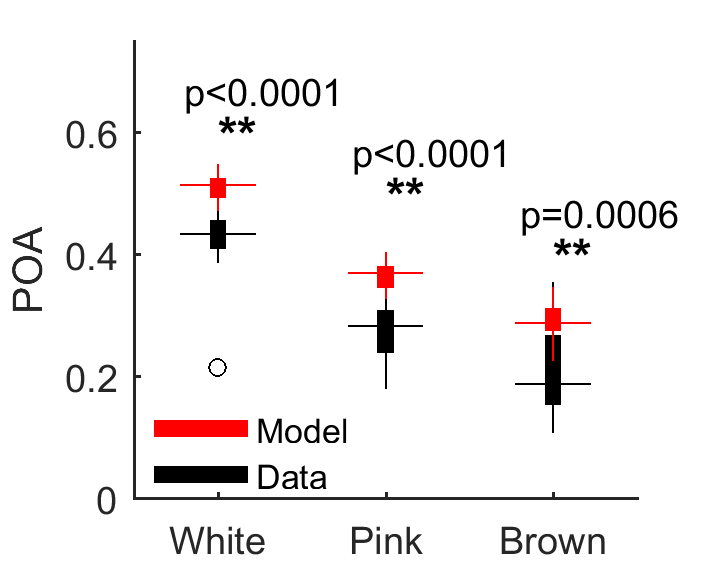}
\caption[\textbf{POA is not sensitive to differences in psychometric curves among stimuli.}]{\textbf{POA is not sensitive to differences in psychometric curves among stimuli.}. In this analysis the average curve over the entire experiment was used for the instantaneous observer. The difference between POA of human observers and those of model observers is still around 0.08 fro all stimulus types.}
\label{fig:POA_sameslope}
\end{figure}

\subsection{Other models tested}
\textit{Sensory-Cognitive model - bias modulations by response only}
\label{subsec:SensoryCognitiveModel2} 
This model is similar to the one presented in the main text, %\ref{subsec:SensoryCognitiveModel1},
only with the adaptation bias regulated by the history of responses rather than of inputs. The output $y$ is filtered with a time constant $\tau$ to give $\overline{y}_i$, which replaced $\overline{x}_i$ in the adaptation variable $A$.
The performance the two models is almost identical.
Note that in this model, the sensory system is regulated by a feedback from the cognitive process, while in the first model this feedback is independent of the cognitive part.
Each of these models is feasible and consistent with our data.
Moreover it is likely that the two loops co-exist.

\begin{figure}[H]
	\centering
	\includegraphics [scale=0.6,trim = 0.0cm 0cm 0cm 0.2cm, clip]{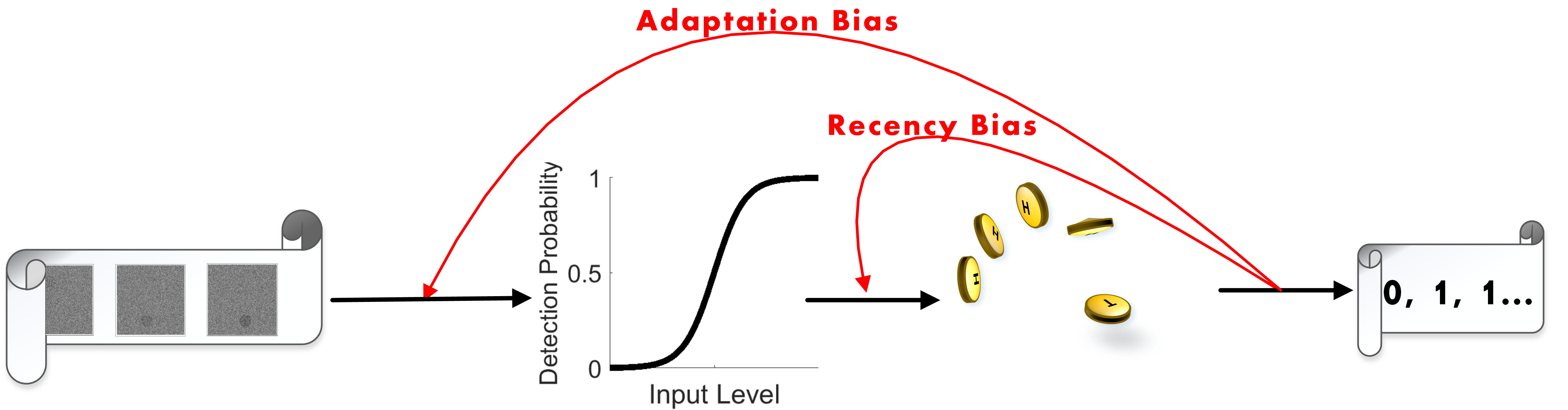}
	\caption[\textbf{Sensory-Cognitive model - dependent only on history of responses}]{\textbf{Sensory - Cognitive model - dependent only on history of responses} - this is almost identical to the model in figure \ref{fig:SensoryCognitiveModel1}, just that the regulation of the adaptation bias of the input is based upon the responses history, rather than on the input history.
		%\includegraphics [scale=0.35,trim = 0.2cm 0.2mm 0cm 1.5cm, clip]{./graphics/Model/modelrevised_ModelIN32.png}
		%\caption[\textbf{Sensory - Cognitive model - dependent only on history of responses}]{\textbf{Sensory - Cognitive model - dependent only on history of responses}  this is almost identical to the model in figure \ref{fig:SensoryCognitiveModel1}, just that the regualtion of the adaptation bias of the input is based upon the diifference of responses histoty for 0.5 detection probability, rather than on the input history distnce from the threshold.
		\label{fig:SensoryCognitiveModel2}}
\end{figure}

%\subsection{Model parameters Setting}
%This part relates to all of the models versions tested.
%The slope and threshold values ($k,\theta$) were based on the averages in experiment.
%The adjustment of other model parameters was done manually in a trial-and-error procedure.
%Although the relationship between parameters in the model was important to each of the results, each parameter by itself regulated mainly a specific feature of the dynamic behavior.
%Specifically: the parameter $\delta P$, positive recency, determines both the POA and the amplitude of the positive phase of the hysteresis curve.
%The parameter \textit{$\tau$} used in the calculation of adaptation bias regulates the strength of hysteresis in the different contextual regimes and the point of zero crossing of $\Delta \theta$ in each of them.
%The parameter of $\gamma$ of the adaptation effect regulates mainly the amplitude of negative hysteresis.

\subsubsection{Cognitive model}
\label{subsec:CognitiveModel}
%\subsubsection{Cognitive model description}
%\label{subsec:CognitiveModelDescription}
In this model the two biases are introduced in the post-sensory, or cognitive, stage of processing.
A significant difference from the first sensory-cognitive combined models is that the calculation of the adaptation biases is dependent only on the history of responses, not on the input history.
\begin{figure}[H]
\centering
\includegraphics[scale=0.6,trim = 0.0cm 0cm 0cm 0.2cm, clip]{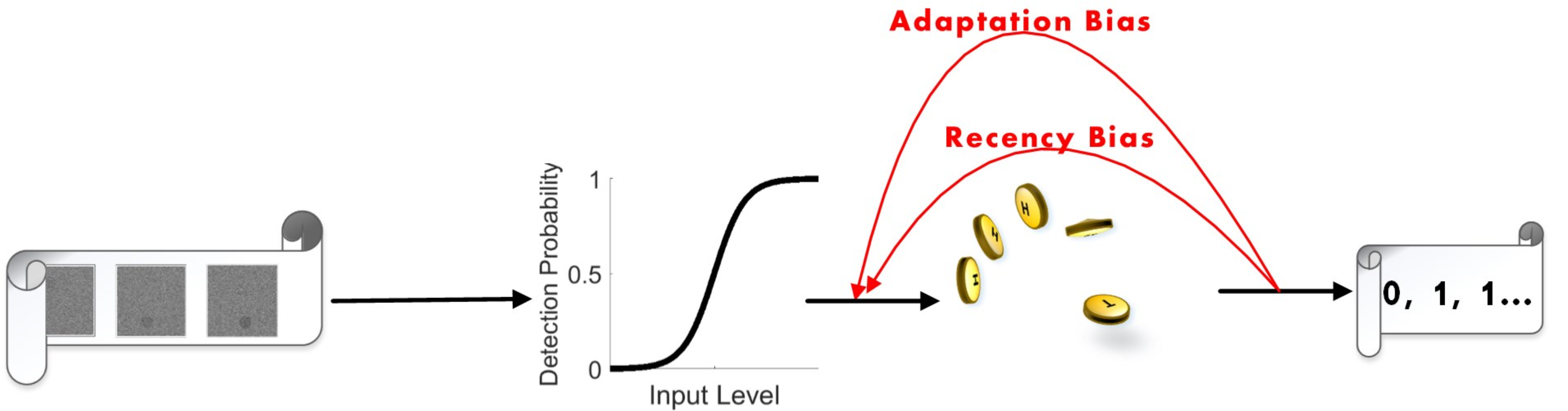}
%\includegraphics [scale=0.35,trim = 0.2cm 0.2mm 0cm 1.2cm, clip]{./graphics/Model/modelrevised_Model9.png}
%{\includegraphics[scale=0.75]{./graphics/Model/ModelGeneral.jpg}}
\caption[\textbf{The cognitive model}]{\textbf{The cognitive model}
Sensory process is instantaneous and independent on history.
It is modeled by constant sigmoidal relations between the momentary input level and the probability of the coin flip.
The cognitive process, on the other side, encapsulates all contextual dependencies.
The probability of the coin flip is a weighting of 3 probabilities the instantaneous with the two biases that depend on the history of responses.
\label{fig:modelAllCognitive}}
\end{figure}

%Analysis of 15 model subjects using the all cognitive model shows that this model is a valid option to describe the data.
\begin{figure}[H]
\label{fig:modelAllCognitive}
\subfloat{\includegraphics[scale=1]{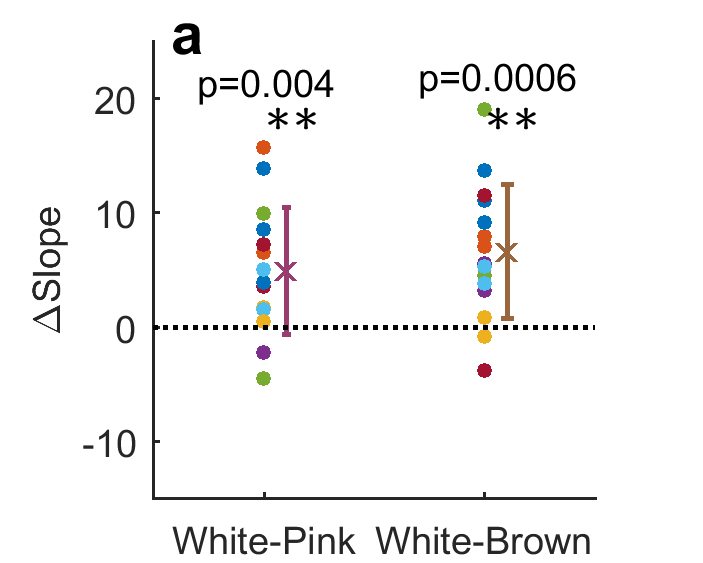}\label{subfig:CognitiveModelSlope}}
\subfloat{\includegraphics[scale=1]{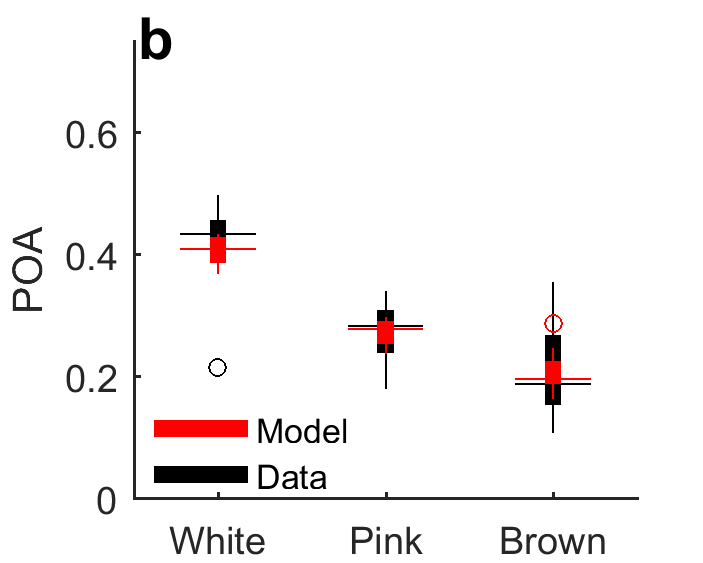}\label{subfig:CognitiveModelPOA}}\\
\subfloat{\includegraphics[scale=1]{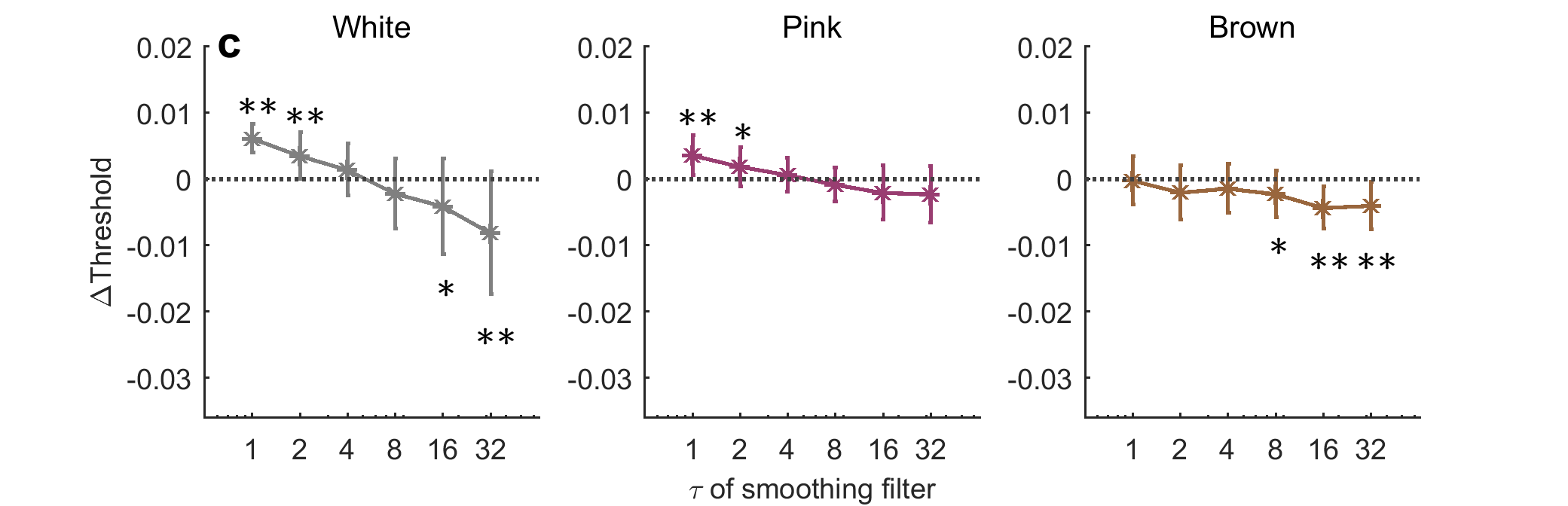}\label{subfig:CognitiveModelHysteresis}}
\caption[\textbf{Cognitive model performance}]{\textbf{Cognitive model Performance}~\protect\subref{subfig:CognitiveModelSlope} - The model show the expected elevation in the slope of \textit{Pink} and \textit{Brown} relative to slope in \textit{White}
~\protect\subref{subfig:CognitiveModelPOA} - the probabilities of alternation in the different modes are reproduced in model subjects.
~\protect\subref{subfig:CognitiveModelHysteresis} - the relation between raising and falling psychometric curves with respects to the different timescales in the model is very similar to the actual data (compare to figure~\ref{subfig:taus}).}
\end{figure}

\end{document}

%% file: structure.tex
\usepackage[
nochapters, % Turn off chapters since this is an article        
beramono, % Use the Bera Mono font for monospaced text (\texttt)
eulermath,% Use the Euler font for mathematics
pdfspacing, % Makes use of pdftex’ letter spacing capabilities via the microtype package
dottedtoc % Dotted lines leading to the page numbers in the table of contents
]{classicthesis} % The layout is based on the Classic Thesis style
\usepackage[T1]{fontenc} % Use 8-bit encoding that has 256 glyphs

\usepackage[utf8]{inputenc} % Required for including letters with accents

\usepackage{graphicx} % Required for including images
\graphicspath{{graphics/}} % Set the default folder for images

\usepackage{enumitem} % Required for manipulating the whitespace between and within lists
\usepackage{xfrac}

\usepackage{lipsum} % Used for inserting dummy 'Lorem ipsum' text into the template
\usepackage{float}
\usepackage{subfig} % Required for creating figures with multiple parts (subfigures)

\usepackage{amsmath,amssymb,amsthm} % For including math equations, theorems, symbols, etc

\usepackage{varioref} % More descriptive referencing
\usepackage[margin=1.2in]{geometry}
\usepackage{breakcites}
\usepackage{setspace} 
\theoremstyle{definition} % Define theorem styles here based on the definition style (used for definitions and examples)

\theoremstyle{plain} % Define theorem styles here based on the plain style (used for theorems, lemmas, propositions)

\theoremstyle{remark} % Define theorem styles here based on the remark style (used for remarks and notes)

\hypersetup{
colorlinks=true, breaklinks=true, bookmarks=true,bookmarksnumbered,
urlcolor=webbrown, linkcolor=RoyalBlue, citecolor=webgreen, % Link colors
pdftitle={}, % PDF title
pdfauthor={\textcopyright}, % PDF Author
pdfsubject={}, % PDF Subject
pdfkeywords={}, % PDF Keywords
pdfcreator={pdfLaTeX}, % PDF Creator
pdfproducer={LaTeX with hyperref and ClassicThesis} % PDF producer
}

%% file: TitlePage.tex
%TitlePage 
%\section*{}
%----------------------------------------------------------------------------------------
%	TITLE AND AUTHOR(S)
%----------------------------------------------------------------------------------------
\title{\normalfont\spacedallcaps{Visual detection of time-varying signals: opposing biases and their timescales}} % The article title
\author{\spacedlowsmallcaps{Urit Gordon\textsuperscript{1,3}, Shimon Marom\textsuperscript{1,3}, Naama Brenner\textsuperscript{2,3}}} 
%\date{2018} % An optional date to appear under the author(s)
%----------------------------------------------------------------------------------------
\maketitle % Print the title/author/date block
\noindent\textit{Number of pages: 33}\newline
\textit{Number of figures: 20}\newline
\textit{Number of words in Abstract: 304, in Introduction: 653, in Discussion: 992}\newline\newline
\textit{Corresponding author: Naama Brenner, nbrenner@technion.ac.il }\newline
\textit{The authors declare no competing financial interests.}\newline
%\textit{The Technion's funding of this research is hereby acknowledged.}\newline
%----------------------------------------------------------------------------------------
%%	AUTHOR AFFILIATIONS
%%----------------------------------------------------------------------------------------
{\let\thefootnote\relax\footnotetext{1 \textit{Faculty of Medicine, Technion, Haifa, 3200003, Israel}}}
{\let\thefootnote\relax\footnotetext{2 \textit{Faculty of Chemical Engineering, Technion, Haifa, 3200003, Israel}}}
{\let\thefootnote\relax\footnotetext{3 \textit{Network Biology Research Lab, Lorry Lockey Interdisciplinary Center for Life Science and Engineering, Technion, Haifa, 3200003, Israel}}}

%% file: Abstract.tex
\section*{Abstract}
% This section will not appear in the table of contents due to the star (\section*)
% 250 words maximum

Human visual perception is a complex, dynamic and fluctuating process. In addition to the incoming visual stimulus, it is affected by many other factors including temporal context, both external and internal to the observer. 
In this study we investigate the dynamic properties of psychophysical responses to a continuous stream of visual near-threshold detection tasks. We manipulate the incoming signals to have temporal structures with various characteristic timescales. Responses of human observers to these signals are analyzed using tools that highlight their dynamical features as well. 

We find that two opposing biases shape perception, and operate over distinct timescales. Positive recency appears over short times, e.g. consecutive trials. Adaptation, entailing an increased probability of changed response, reflects trends over longer times. Analysis of psychometric curves conditioned on various temporal events reveals that the balance between the two biases can shift depending on their interplay with the temporal properties of the input signal.  
A simple mathematical model reproduces the experimental data in all stimulus regimes. Taken together, our results
support the view that visual response fluctuations reflect complex internal dynamics, possibly related to higher cognitive processes.

%% file: AuthorSummary.tex
\section{Author summary}
%{Significance Statement}
%120 words max
\paragraph{}
Human perception exhibits internal dynamics and fluctuates even when external conditions are well controlled. In this study we take the dynamic approach to study these fluctuations, manipulating temporal properties of incoming stream of signals and analyzing dynamic aspects of the responses. Such an approach can reveal internal biases in perception and their characteristic timescales.

We find that human observers exhibit two opposing biases: in the short term they tend to repeat their response, while in the long term they tend to change it. The balance between the two depends on the incoming signal stream. This behavior can be interpreted as an exploration-exploitation trade-off; it suggests that dynamics of perception reflects higher cognitive processes in addition signal detection itself. 

%% file: Article1_09_04.bbl
\begin{thebibliography}{}

\bibitem[Abrahamyan et~al., 2016]{Abrahamyan2016a}
Abrahamyan, A., Silva, L.~L., Dakin, S.~C., Carandini, M., and Gardner, J.~L.
  (2016).
\newblock {Adaptable history biases in human perceptual decisions.}
\newblock {\em Proc. Natl. Acad. Sci.}, 113(25):E3548--E3557.

\bibitem[Anderson, 1971]{Anderson1971}
Anderson, H.~N. (1971).
\newblock {Test of adaptation-level theory as an explanation of a recency
  effect in psychophysical integration}.
\newblock {\em J. Exp. Psychol.}, 87(1):57--63.

\bibitem[Barack and Gold, 2016]{Barack2016}
Barack, D.~L. and Gold, J.~I. (2016).
\newblock {Temporal trade-offs in psychophysics}.
\newblock {\em Curr. Opin. Neurobiol.}, 37:121--125.

\bibitem[Benda and Herz, 2003]{BendaHerz2003}
Benda, J. and Herz, A. V.~M. (2003).
\newblock {A Universal Model for Spike-Frequency Adaptation}.
\newblock {\em Neural Comput.}, 15(11):2523----2564.

\bibitem[Blackwell, 1952]{Blackwell1952}
Blackwell, H.~R. (1952).
\newblock {Studies of Psychophysical Methods for Measuring Visual Thresholds}.
\newblock {\em J. Opt. Soc. Am.}, 42(9):606.

\bibitem[Brenner et~al., 2000]{Brenner2000}
Brenner, N., Bialek, W., and {de Ruyter van Steveninck}, R. (2000).
\newblock {Adaptive Rescaling Maximizes Information Transmission}.
\newblock {\em Neuron}, 26(3):695--702.

\bibitem[Chopin and Mamassian, 2012]{Chopin2012a}
Chopin, A. and Mamassian, P. (2012).
\newblock {Predictive Properties of Visual Adaptation}.

\bibitem[Cross, 1973]{Cross1973}
Cross, D.~V. (1973).
\newblock {Sequential dependencies and regression in psychophysical judgments}.
\newblock {\em Percept. Psychophys.}, 14(3):547--552.

\bibitem[{Duncan Luce} et~al., 1982]{Luce1982}
{Duncan Luce}, R., Nosofsky, R.~M., Green, D.~M., and Smith, A.~F. (1982).
\newblock {The bow and sequential effects in absolute identification}.
\newblock {\em Percept. Psychophys.}, 32(5):397--408.

\bibitem[Faisal et~al., 2008]{Faisal2008}
Faisal, A.~A., Selen, L. P.~J., and Wolpert, D.~M. (2008).
\newblock {Noise in the nervous system.}
\newblock {\em Nat. Rev. Neurosci.}, 9(april):292--303.

\bibitem[Fischer and Whitney, 2014]{Fischer2014}
Fischer, J. and Whitney, D. (2014).
\newblock {Serial dependence in the perception of faces}.
\newblock {\em Curr. Biol.}, 24(21):2569--2574.

\bibitem[Freiberg, 1937]{Freiberg1937}
Freiberg, A.~D. (1937).
\newblock {'Fluctuations of Attention' with Weak Tactual Stimuli: A Study in
  Perceiving}.
\newblock {\em Am. J. Psychol.}, 49(1):23--36.

\bibitem[Fr{\"{u}}nd et~al., 2014]{Frund2014}
Fr{\"{u}}nd, I., Wichmann, F.~A., and Macke, J.~H. (2014).
\newblock {Quantifying the effect of intertrial dependence on perceptual
  decisions.}
\newblock {\em J. Vis.}, 14(7):1--16.

\bibitem[Gepshtein and Kubovy, 2005]{Gepshtein2005}
Gepshtein, S. and Kubovy, M. (2005).
\newblock {Stability and change in perception: spatial organization in temporal
  context}.
\newblock {\em Exp. Brain Res.}, 160(4):487--495.

\bibitem[Gibson and Radner, 1937]{Gibson1937}
Gibson, J.~J. and Radner, M. (1937).
\newblock {Adaptation, after-effect and contrast in the perception of tilted
  lines. I. Quantitative studies.}
\newblock {\em J. Exp. Psychol.}, 20(5):453--467.

\bibitem[Holland and Lockhead, 1968]{Holland1968}
Holland, M.~K. and Lockhead, G.~R. (1968).
\newblock {Sequential effects in absolute judgments of loudness}.
\newblock {\em Percept. {\{}{\&}{\}} Psychophys.}, 3(6):409--414.

\bibitem[Howarth and Bulmer, 1956]{Howarth1956}
Howarth, C.~I. and Bulmer, M.~G. (1956).
\newblock {Non-random sequences in visual threshold experiments}.
\newblock {\em Q. J. Exp. Psychol.}, 8(4):163--171.

\bibitem[Laughlin, 1981]{Laughlin1981}
Laughlin, S. (1981).
\newblock {A simple coding procedure enhances a neuron's information capacity}.

\bibitem[Lockhead, 1970]{Lockhead1970}
Lockhead, G.~R. (1970).
\newblock {Identification and the form of multidimensional discrimination
  space.}
\newblock {\em J. Exp. Psychol.}, 85(1):1--10.

\bibitem[Magnussen and Greenlee, 1999]{Magnussen1999}
Magnussen, S. and Greenlee, M.~W. (1999).
\newblock {The psychophysics of perceptual memory}.
\newblock {\em Psychol. Res.}, 62(2):81--92.

\bibitem[Mcgill, 1957]{Mcgill1957}
Mcgill, W.~J. (1957).
\newblock {Serial effects in auditory threshold judgments.}
\newblock {\em J. Exp. Psychol.}, 5(53):297.

\bibitem[Parducci, 1964]{Parducci1964}
Parducci, A. (1964).
\newblock {Sequential effects in judgment.}
\newblock {\em Psychol. Bull.}, 61(3):163--167.

\bibitem[Parducci and Sandusky, 1965]{Parducci1965}
Parducci, A. and Sandusky, A. (1965).
\newblock {Distribution and sequence effects in judgment}.
\newblock {\em J. Exp. Psychol.}, 69(5):450--459.

\bibitem[Pollack, 1954]{Pollack1954}
Pollack, I. (1954).
\newblock {Intensity Discrimination Thresholds under Several Psychophysical
  Procedures}.
\newblock {\em J. Acoust. Soc. Am.}, 26(6):1056.

\bibitem[Raviv et~al., 2012]{Raviv2012}
Raviv, O., Ahissar, M., and Loewenstein, Y. (2012).
\newblock {How Recent History Affects Perception: The Normative Approach and
  Its Heuristic Approximation}.
\newblock {\em PLoS Comput. Biol.}, 8(10).

\bibitem[Sanabria et~al., 2004]{Sanabria2004}
Sanabria, D., ngel Correa, Lupi~ez, J., Spence, C., Correa, A.,
  Lupi{\'{a}}{\~{n}}ez, J., and Spence, C. (2004).
\newblock {Bouncing or streaming? Exploring the influence of auditory cues on
  the interpretation of ambiguous visual motion}.
\newblock {\em Exp. Brain Res.}, 157(4):537--41.

\bibitem[Schwiedrzik et~al., 2014]{Schwiedrzik2014}
Schwiedrzik, C.~M., Ruff, C.~C., Lazar, A., Leitner, F.~C., Singer, W., and
  Melloni, L. (2014).
\newblock {Untangling perceptual memory: hysteresis and adaptation map into
  separate cortical networks.}
\newblock {\em Cereb. cortex}, 24(5):1152--64.

\bibitem[Snyder et~al., 2015]{Snyder2015}
Snyder, J.~S., Schwiedrzik, C.~M., Vitela, A.~D., and Melloni, L. (2015).
\newblock {How previous experience shapes perception in different sensory
  modalities.}
\newblock {\em Front. Hum. Neurosci.}, 9:594.

\bibitem[Verplanck et~al., 1952]{Verplanck1952}
Verplanck, W.~S., Collier, G.~H., and Cotton, J.~W. (1952).
\newblock {Nonindependence of successive responses in measurements of the
  visual threshold}.
\newblock {\em J. Exp. Psychol. , Am. Psychol. Assoc.}, 44(4):273.

\bibitem[Ward, 1973]{Ward1973}
Ward, L.~M. (1973).
\newblock {Repeated magnitude estimations with a variable standard: Sequential
  effects and other properties}.
\newblock {\em Percept. {\{}{\&}{\}} Psychophys.}, 13(2):193--200.

\bibitem[Wu et~al., 2009]{Wu2009}
Wu, J., Xu, H., Dayan, P., and Qian, N. (2009).
\newblock {The Role of Background Statistics in Face Adaptation}.
\newblock {\em J. Neurosci.}, 29(39):12035--12044.

\end{thebibliography}
